\documentclass{aa}  

\usepackage{graphicx}
\usepackage{txfonts}
\usepackage{bm}	
\usepackage[table]{xcolor}
\definecolor{LightGray}{gray}{0.9}
\usepackage{booktabs}
\usepackage{fontawesome}
\usepackage{soul}
\usepackage{tcolorbox}

\usepackage{listings}

\definecolor{codegreen}{rgb}{0,0.6,0}
\definecolor{codegray}{rgb}{0.5,0.5,0.5}
\definecolor{codepurple}{rgb}{0.58,0,0.82}
\definecolor{backcolour}{rgb}{0.95,0.95,0.92}
 
\lstdefinestyle{mystyle}{
    backgroundcolor=\color{backcolour},   
    commentstyle=\color{codegreen},
    keywordstyle=\color{magenta},
    numberstyle=\tiny\color{codegray},
    stringstyle=\color{codepurple},
    basicstyle=\footnotesize,
    breakatwhitespace=false,         
    breaklines=true,                 
    captionpos=b,                    
    keepspaces=true,                 
    numbers=left,                    
    numbersep=5pt,                  
    showspaces=false,                
    showstringspaces=false,
    showtabs=false,                  
    tabsize=2
}

\usepackage[pdfencoding=auto,psdextra]{hyperref}
\hypersetup{
    colorlinks=true,
    linkcolor=blue,
    filecolor=magenta,      
    urlcolor=blue,
    citecolor=blue
}
\newcommand{\dd}{\mathrm{d}}
\newcommand{\e}[1]{_{\text{#1}}}
\newcommand{\ev}[1]{\left\langle #1 \right\rangle}

\newcommand{\pysco}{\texttt{PySCo}\xspace}
\newcommand{\ecosmog}{\texttt{ECOSMOG}\xspace}
\newcommand{\emantis}{\texttt{e-MANTIS}\xspace}
\newcommand{\ramses}{\texttt{RAMSES}\xspace}
\newcommand{\numba}{\texttt{Numba}\xspace}

\begin{document}
\lstset{
    frame       = single,
    numbers     = left,
    showspaces  = false,
    showstringspaces    = false,
    captionpos  = t,
}

   \title{PySCo: A fast particle-mesh $N$-body code for modified gravity simulations in Python}

   \author{Michel-Andrès Breton}

   \institute{
   Université Paris-Saclay, Université Paris Cité, CEA, CNRS, AIM, 91191, Gif-sur-Yvette, France
   \and
   Institute of Space Sciences (ICE, CSIC), Campus UAB, Carrer de
            Can Magrans, s/n, 08193 Barcelona, Spain
    \and 
    Laboratoire Univers et Théorie, Observatoire de Paris, Université
            PSL, Université Paris Cité, CNRS, 92190 Meudon, France \\
             \email{michel-andres.breton@cea.fr}
             }


 
  \abstract{
We present \pysco, a fast and user-friendly Python library designed to run cosmological $N$-body simulations across various cosmological models, such as $\Lambda$CDM ($\Lambda$ with cold dark matter) and $w_0w_a$CDM, and alternative theories of gravity, including $f(R)$, MOND (modified newtonian dynamics) and time-dependent gravitational constant parameterisations. \pysco employs particle-mesh solvers, using multigrid or fast Fourier transform (FFT) methods in their different variations. Additionally, \pysco can be easily integrated as an external library, providing utilities for particle and mesh computations.
The library offers key features, including an initial condition generator based on up to third-order Lagrangian perturbation theory (LPT), power spectrum estimation, and computes the background and growth of density perturbations. In this paper, we detail \pysco's architecture and algorithms and conduct extensive comparisons with other codes and numerical methods. Our analysis shows that, with sufficient small-scale resolution, the power spectrum at redshift $z = 0$ remains independent of the initial redshift at the 0.1\% level for $z_{\rm ini} \geq$ 125, 30, and 10 when using first, second, and third-order LPT, respectively. Moreover, we demonstrate that acceleration (or force) calculations should employ a configuration-space finite-difference stencil for central derivatives with at least five points, as three-point derivatives result in significant power suppression at small scales. Although the seven-point Laplacian method used in multigrid also leads to power suppression on small scales, this effect can largely be mitigated when computing ratios.
In terms of performance, \pysco only requires approximately one CPU hour to complete a Newtonian simulation with $512^3$ particles (and an equal number of cells) on a laptop. Due to its speed and ease of use, \pysco is ideal for rapidly generating vast ensemble of simulations and exploring parameter spaces, allowing variations in gravity theories, dark energy models, and numerical approaches. This versatility makes \pysco a valuable tool for producing emulators, covariance matrices, or training datasets for machine learning.
  }

   \keywords{large-scale structure of Universe -- Gravitation -- Methods: numerical -- Cosmology: miscellaneous}

   \maketitle
%

\section{Introduction}

On the largest scales, the $\Lambda$CDM ($\Lambda$ with cold dark matter) model provides a robust description of our Universe, where dark energy is represented as a cosmological constant ($\Lambda$) and dark matter as non-relativistic cold dark matter \citep{planck2018cosmological,alam2021completed,riess2022comprehensive}. The $\Lambda$CDM model also assumes General Relativity (GR), which offers an accurate representation of the Universe by incorporating small perturbations on a Friedman-Lemaître-Robertson-Walker (FLRW) background \citep{green2014how}. The formation of the large-scale structure is driven by these small perturbations, which grow over time through gravitational interactions, giving rise to the cosmic web observed today \citep{peebles1980LSS}.

While linear theory can analytically describe the growth of perturbations on large scales above $\sim$100 Mpc \citep{yoo2009new, bonvin2011what, challinor2011linear}, the non-linear nature of structure formation on smaller scales necessitates more advanced approaches. Higher-order perturbation theory, both in its Eulerian \citep{peebles1980LSS, bernardeau2002large} and Lagrangian \citep{zeldovich1970gravitational, buchert1993lagrangian} formulations, allows for a more accurate understanding down to scales of $\sim$20 Mpc. Below this scale, structure formation becomes highly non-linear, and the evolution of gravitationally interacting particles can only be accurately modeled through $N$-body simulations \citep{efstathiou1985numerical}.

Over time, numerous cosmological $N$-body codes have been developed \citep{couchman1991mesh, kravtsov1997ART, knebe2001MLAPM, teyssier2002cosmological, ishiyama2009greeM, potter2017pkdgrav3, springel2021simulating, garrison2021abacus}, with recent simulations including trillions of particles \citep{ishiyama2021uchuu, castander2024fs2}. While significant effort has been invested in developing efficient Newtonian simulations for cosmology (see \citealt{angulo2022large} for a review), many $N$-body codes for alternative gravity theories are derived from Newtonian codes. For instance, \texttt{ECOSMOG} \citep{li2012ecosmog}, \texttt{ISIS} \citep{llinares2014isis}, \texttt{RayMOND} \citep{candlish2015raymond} , and \texttt{Phantom of RAMSES} \citep{lughausen2015por} are all based on the \texttt{RAMSES} code \citep{teyssier2002cosmological}. The former two implement the $f(R)$ model \citep{hu2007models} and the nDGP model \citep{dvali20004Dgravity}, while the latter two implement MOND (modified newtonian dynamics, \citealt{milgrom1983mond}). Although it is possible to integrate modified gravity theories into TreePM codes like \texttt{Gadget} \citep{springel2021simulating}, as demonstrated by \texttt{MG-Gadget} \citep{puchwein2013mggadget}, \texttt{RAMSES} has emerged as the preferred code for such implementations.
This is due to the fact that alternative theories of gravity generally introduce additional fields governed by non-linear partial differential equations, which cannot be efficiently solved using standard tree-based methods. As a result, \texttt{RAMSES}, a particle-mesh (PM) code with adaptive-mesh refinement (AMR) and a multigrid solver, becomes an ideal choice for implementing these features. However, PM-AMR, TreePM, or Fast Multipole Method (FMM) codes typically require significant computational resources, even when highly optimised. This challenge is exacerbated in simulations of alternative gravity theories; for instance, \cite{adamek2024KPJC6P1} showed that $f(R)$ or nDGP simulations can run up to ten times slower than their Newtonian counterparts.

To address these computational demands, researchers have explored PM codes \citep{knebe2001MLAPM, merz2005towards, feng2016fastpm, adamek2016gevolution, klypin2018dark} to reduce the cost at the expense of small-scale accuracy. These faster, albeit less precise, simulations are suitable for specific applications, such as producing large numbers of realisations for covariance matrix estimation or calculating the ratio (or boost) of specific statistical quantities relative to a reference case exploring large parameter spaces. 
Furthermore, exact PM codes can accurately reproduce structure formation at small scales given a higher resolution of the uniform mesh. Consequently, such codes have been ideal for developing solvers for modified gravity theories (see \citealt{llinares2018simulation} for a review), and have been used to study their impact relative to Newtonian simulations \citep{valogiannis2016efficient, winther2017mgcola, hassani2020nbody, ruan2022fast, hernandezaguayo2022fast}.

Given the need for speed, most $N$-body simulations have traditionally been written in compiled languages such as Fortran, C, or C\texttt{++}. In contrast, Python has become the most popular programming language in data science, owing to its straightforward syntax, rapid development speed, and extensive community-driven libraries, especially in astronomy. This popularity has created a gap between simulators and the broader scientific field. Despite its advantages, Python is often viewed as a slow language when used natively. To address this, significant effort has gone into developing efficient Python libraries, either as wrappers for C-based codes (such as \texttt{NumPy}, \citealt{harris2020array}) or through compiling Python code to machine language, as seen in \texttt{Numba} \citep{lam2015numba} and \texttt{Cython} \citep{behnel2011cython}. Recently, the latter approach was used to develop the P$^3$M code CO$N$CEPT \citep{dakin2022cosmological}, demonstrating the viability of Python for high-performance applications.

In this paper, we present \pysco\footnote{\faicon{github}~\href{https://github.com/mianbreton/pysco}{https://github.com/mianbreton/pysco}} (Python Simulations for Cosmology), a cosmological PM $N$-body code written in Python and utilising the \numba library for increased performance and multithreading. The paper is organised as follows: Section~\ref{sec:theory} introduces the different models implemented in \pysco, including the modified gravity theories $f(R)$ from \cite{hu2007models}, MOND \citep{milgrom1983mond}, parameterised gravity \citep{amendola2008measuring}, and dynamic dark energy \citep{chevallier2001accelerating, linder2003exploring}. Section~\ref{sec:methods} details the structure and algorithms implemented in the code, covering initial condition generation and various $N$-body solvers. In Section~\ref{sec:results}, we validate \pysco against other codes and analyse the impact of different numerical methods on the matter power spectrum. Finally, we conclude in Section~\ref{sec:conclusions}.
\section{Theory}
\label{sec:theory}
\subsection{Newtonian gravity}

Let us consider only scalar perturbations on FLRW metric, in the Newtonian gauge \citep{ma1995cosmological}
\begin{equation}
\dd s^2 = a^2(\eta)\left[-\left(1+2\psi\right)\dd\eta^2 + \left(1-2\phi\right)\dd\bm{x}^2\right],
\label{eq:perturbed_FLRW_metric}
\end{equation}
with $a$ the scale factor, $\eta$ the conformal time and $\psi$ and $\phi$ the Bardeen potentials \citep{bardeen1980gauge}. In GR, we have $\psi = \phi$, and the Einstein equation gives
\begin{equation}
\label{eq:einstein_gr}
    \nabla^2\phi -3\mathcal{H}\left(\phi' + \mathcal{H}\phi\right) = 4\pi Ga^2 \delta\rho,
\end{equation}
with $\mathcal{H} = aH$ the conformal Hubble parameter, $\phi'$ the time derivative of the potential and $\rho$ the Universe's components density. It is common to apply the Newtonian and quasi-static (neglecting time derivatives) approximations, which involve neglecting the second term on the left-hand side of Eq.~\eqref{eq:einstein_gr}, as it is only significant at horizon scales. In the context of Newtonian cosmology, the Einstein equation takes the same form as the classical Poisson equation, with an additional dependence on the scale factor
\begin{equation}
    \nabla^2\phi = 4\pi G a^2\delta\rho_m,
\end{equation}
with $\rho_m$ the matter density.
There are, however, cosmological General Relativity (GR) simulations \citep{adamek2016gevolution, barrera-hinojosa2020gramses} that solve the full Einstein equations in the weak-field limit, taking into account gauge issues \citep{fidler2015general, fidler2016relativistic}. Additionally, there are methods to interpret Newtonian simulations within a relativistic framework \citep{chisari2011connection, adamek2019large}. In this work, however, we focus on smaller scales where relativistic effects are negligible, and the Newtonian approximation remains well justified.

\subsection{Dynamical dark energy}

The $w_0w_a$CDM model provides a useful phenomenological extension of the standard $\Lambda$CDM framework by offering a dynamic, time-dependent description of dark energy. In this model, the cosmological constant is replaced by a variable dark energy component, which affects the formation of cosmic structures by modifying the universe's expansion history. For a flat geometry ($\Omega_k = 0$), the Hubble parameter $H(z)$ is given by
\begin{equation}
    H(z) = H_0 \sqrt{\Omega_{m,0} (1+z)^3 + \Omega_{r,0} (1+z)^4 + \Omega_\Lambda(z)},
\end{equation}
where a subscript zero indicates a present-day evaluation, and $\Omega_m$, $\Omega_r$, and $\Omega_\Lambda$ represent the density fractions of matter, radiation, and dark energy, respectively. In this model, the dark-energy density evolves with redshift, following the relation
\begin{equation}
    \Omega_\Lambda(z) = \Omega_{\Lambda, 0} \exp{\left\{\int_0^z\frac{3\left[1+w(z')\right]\dd z'}{1+z'} \right\}}.
\end{equation}
Using the widely adopted CPL parametrisation \citep{chevallier2001accelerating, linder2003exploring}, the dark-energy equation of state is expressed as
\begin{equation}
    w(z) = w_0 + w_a \frac{z}{1+z}.
\end{equation}
This simple modification does not alter the Einstein field equations nor the equations of motion, and it recovers the standard $\Lambda$CDM model when $w_0 = -1$ and $w_a = 0$.

\subsection{MOND gravity}
\label{sec:theory_mond}
The Modified Newtonian Dynamics (MOND) theory, introduced by \cite{milgrom1983mond}, was proposed as a potential solution to the dark matter problem by suggesting a deviation from Newtonian gravity in a Universe where the matter content is entirely in the form of baryons (for a detailed review, see \citealt{famaey2012mond}). In MOND, Newton’s second law is modified as follows
\begin{equation}
    \mu\left(\frac{|\bm{g}|}{g_0}\right) \bm{g} = \bm{g}_N,
\end{equation}
where $\bm{g}$ and $\bm{g}_N$ represent the MOND and Newtonian accelerations, respectively, and $g_0$ is a characteristic acceleration scale, approximately $g_0 \approx cH_0/[2\pi] \approx 10^{-10}$m.s$^{-2}$. The function $\mu(x)$ is an interpolating function that governs the transition between the Newtonian regime (where gravitational force scales as $r^{-2}$) and the MOND regime (where the force scales as $r^{-1}$), with $r$ being the separation between masses. The interpolating function has the following limits
\begin{equation}
\mu(x) = 
\left\lbrace
\begin{array}{ll}
1, &  x \gg 1, \\
x, &  x \ll 1. \\
\end{array}\right.
\end{equation}
Similarly, the inverse interpolating function $\nu(y)$ can be defined as
\begin{equation}
    \bm{g} = \nu\left(\frac{\bm{g}_N}{g_0}\right) \bm{g}_N,
\end{equation}
where $\nu(y)$ follows the limits
\begin{equation}
\nu(y) = 
\left\lbrace
\begin{array}{ll}
1, &  y \gg 1, \\
y^{-1/2}, &  y \ll 1. \\
\end{array}\right.
\end{equation}
In the MOND framework, the classical Poisson equation is modified as follows \citep{bekenstein1984does}
\begin{equation}
\label{eq:MOND_AQUAL_poisson}
    \nabla\left[\mu\left(\frac{|\nabla\phi|}{g_0}\right)\nabla\phi\right] = 4\pi G\delta\rho,
\end{equation}
a formulation known as AQUAL, derived from a quadratic Lagrangian. In this paper, however, we consider the QUMOND (quasi-linear MOND, \citealt{milgrom2010quasilinear}) formulation, where the non-linearity in the Poisson equation is re-expressed in terms of an additional effective dark matter fluid in the source term. The modified Poisson equations are
\begin{align}
\label{eq:QUMOND_Poisson_Newton}
&\nabla^2\phi^{\rm N} = 4\pi G\rho, \\
\label{eq:QUMOND_Poisson_MOND}
&\nabla^2\phi = \nabla\left[\nu\left(\left|\nabla\phi^{\rm N}\right|/g_0\right)\nabla\phi^{\rm N}\right],
\end{align}
where $\phi^N$ and $\phi$ are the Newtonian and MOND potentials, respectively. This system of equations is more convenient to solve, as it involves only two linear Poisson equations. Moreover, numerical simulations have demonstrated that the AQUAL and QUMOND formulations yield very similar results \citep{candlish2015raymond}.

The missing component in the MOND framework is the specific form of the $\nu(y)$ function. Several families of interpolating functions have been proposed, each with different characteristics

\textbf{- Simple function:} 
\begin{equation}
\label{eq:mond_simple}
    \nu(y) = \frac{1}{2} + \frac{\sqrt{1 + 4/y}}{2},
\end{equation}
which corresponds to the simple function proposed by \cite{famaey2005mond}, equivalent to $\mu(x) = x/(1+x)$. 

\textbf{- The $n$-family:}
\begin{equation}
    \nu(y) = \left[\frac{1}{2} + \frac{\sqrt{1 + 4/y^n}}{2}\right]^{1/n},
\end{equation}
a commonly used parametrisation for the interpolating function \citep{milgrom2008rings}. The case $n = 2$ is particularly well studied and is known as the standard interpolating function \citep{begeman1991extended}.
Additionally, \cite{milgrom2008rings} introduced other functional forms

\textbf{- The $\beta$-family:}
\begin{equation}
    \nu(y) = \left(1 - e^{-y}\right)^{-1/2} + \beta e^{-y},
\end{equation}

\textbf{- The $\gamma$-family:}
\begin{equation}
    \nu(y) = \left(1 - e^{-y^{\gamma/2}}\right)^{-1/\gamma} + \left(1-\gamma^{-1}\right)e^{-y^{\gamma/2}},
\end{equation}

\textbf{- The $\delta$-family:}
\begin{equation}
    \nu(y) = \left(1 - e^{-y^{\delta/2}}\right)^{-1/\delta} ,
\end{equation}
which is a subset of the $\gamma$-family.
While we focus on non-relativistic formulations of MOND for simplicity, it is important to acknowledge that various relativistic frameworks have been developed \citep{bekenstein2006mond, milgrom2010bimetric, skordis2021new}, along with a recent generalisation of QUMOND \citep{milgrom2023gqumond}. These topics, however, are beyond the scope of this paper.

\subsection{Parametrised gravity}
\label{sec:theory_parametrized_gravity}

A straightforward and effective phenomenological approach to modifying the theory of gravity is through the $\mu-\Sigma$ parametrisation of the Einstein equations \citep{amendola2008measuring}. This is particularly useful when considering unequal Bardeen potentials $\phi \neq \psi$. Under the Newtonian and quasi-static approximations, the Einstein equations can be expressed as
\begin{align}
\label{eq:parametrized_newtonian_equation}
    \nabla^2\psi = 4\pi G \mu(a)a^2\delta\rho_m, \\
    \nabla^2 (\psi + \phi) = 8\pi G\Sigma(a)a^2\delta\rho_m,
\end{align}
where  $\mu(a)$ represents the time-dependent `effective gravitational coupling', which can be interpreted as a modification of the gravitational constant, and $\Sigma(a)$ is the `light deflection parameter'. Since our focus is on the evolution of dark-matter particles, we only need to implement Eq.~\eqref{eq:parametrized_newtonian_equation}, which involves $\mu(a)$.
In practice, the gravitational coupling $\mu(a)$ could be a function of both time and scale in Fourier space, $\mu(a,k)$, as in the `effective-field theory of dark energy' \citep{frusciante2020eftofde}. However, for simplicity, we prefer methods that can be solved numerically in both Fourier and configuration space. The inclusion of scale-dependent corrections will be considered in future work.

For the functional form of $\mu(a)$, we use the parametrisation from \cite{simpson2013cfhtlens, planck2016de_mg, planck2018cosmological,abbott2019des1yrextended}, which allows for deviations from GR during a dark-energy dominated era
\begin{equation}
    \label{eq:mu_parametrized_gravity}
    \mu(a) = 1 + \mu_0 \frac{\Omega_\Lambda(a)}{\Omega_{\Lambda, 0}},
\end{equation}
where $\mu_0$ is the only free parameter, representing the gravitational coupling today.

\subsection{$f(R)$ gravity}
\label{sec:theory_fr}

In $f(R)$ gravity, the Lagrangian extends the Einstein-Hilbert action (GR) by including an arbitrary function of the Ricci scalar curvature $R$. The total action is given as
\citep{buchdal1970nonlinear,sotiriou2010reviews}
\begin{equation}
    S = \int \dd^4x \sqrt{-g} \left[ \frac{R + f(R)}{16\pi G} + \mathcal{L}_m\right],
\end{equation}
where $\mathcal{L}_m$ represents the matter Lagrangian, $G$ is the gravitational constant, $g$ is the determinant of the metric, and $f(R)$ is the additional function of the curvature, which reduces to $-2\Lambda$  in the standard $\Lambda$CDM model. A commonly used parametrisation of $f(R)$ gravity is provided by  \cite{hu2007models}, with the following functional form
\begin{align}
    f(R) &= -m^2\frac{c_1\left(R/m^2\right)^n}{c_2\left(R/m^2\right)^n + 1}, \\
     &\approx -\frac{c_1}{c_2}m^2 + \frac{c_1}{c_2^2}m^2 \left(\frac{m^2}{R}\right)^n, \quad R\gg m^2,
\end{align}
where $n, c_1$ and $c_2$ are the model parameters, and $m$ represents the curvature scale, given by $m^2 = \Omega_m H_0^2/c^2 = 8\pi G\bar{\rho}_{m,0}/(3c^2)$, where $\bar{\rho}_{m,0}$ is the current mean matter density. This model incorporates a Chameleon screening mechanism \citep{khoury2004chameleon,burrage2018tests} to suppress the fifth force  caused by the scalar field $f_R$ (also known as the `scalaron'). The scalaron is given by
\begin{equation}
    f_R = \frac{\dd f(R)}{\dd R} \approx -n\frac{c_1}{c_2^2}\left(\frac{m^2}{R}\right)^{n+1},
\end{equation}
which allows the theory to recover GR in high-density environments, ensuring consistency with Solar System tests. Observational evidence for dark energy as a cosmological constant imposes the constraint
\begin{equation}
    \frac{c_1}{c_2} = 6\frac{\Omega_{\Lambda, 0}}{\Omega_{m, 0}},
\end{equation}
and we can also express
\begin{equation}
    \frac{c_1}{c_2^2} = -\frac{3}{n} \left[1+4\frac{\Omega_{\Lambda, 0}}{\Omega_{m, 0}} \right]^{n+1} f_{R0},
\end{equation}
where $f_{R0}$ is the present-day value of the scalaron, with current observational constraints from galaxy clusters indicating $\log_{10}f_{R0} < -5.32$ \citep{vogt2025constraints}. In this framework, the Poisson equation is modified compared to its Newtonian counterpart. There is an additional term that depends on the scalaron field
\begin{eqnarray}
\label{eq:fr_theory_1}
    \nabla^2\phi &=& \frac{16\pi G a^2}{3} \delta\rho - \frac{1}{6}\delta R, \\
\label{eq:fr_theory_2}
    \nabla^2 f_R &=& -\frac{8\pi Ga^2}{3c^2} \delta\rho + \frac{1}{3}\delta R,
\end{eqnarray}
where the difference in curvature is given by
\begin{equation}
    \delta R = R - \bar{R} = \bar{R}\left[\left(\frac{\bar{f}_{R}}{f_R}\right)^{1/(n+1)} - 1\right].
\end{equation}
Here, $\bar{R}$ represents the background curvature, expressed as
\begin{equation}
    \bar{R} = 3m^2\left(a^{-3} + 4\frac{\Omega_\Lambda}{\Omega_m}\right),
\end{equation}
and $\bar{f}_R$ is the background value of $f_R$
\begin{equation}
    \bar{f}_R = \left(\frac{\bar{R}_0}{\bar{R}}\right)^{n+1}f_{R,0}.
\end{equation}
The primary observational distinction between $f(R)$ gravity and GR lies in the enhanced clustering on small scales, with the amplitude and shape of these deviations being dependent on $f_{R0}$, despite the two models sharing the same overall expansion history.

\section{Methods}
\label{sec:methods}
This section reviews the numerical methods used in \pysco, from generating initial conditions to evolving dark-matter particles in $N$-body simulations across various theories of gravity.

\pysco is entirely written in Python and uses the open-source library \texttt{Numba} \citep{lam2015numba}, which compiles Python code into machine code using the LLVM compiler. This setup combines Python's high development speed and rich ecosystem with the performance of C/Fortran. To optimise performance, \pysco relies on writing native Python code with `for' loops, similar to how it would be done in C or Fortran.

\texttt{Numba} integrates seamlessly with \texttt{NumPy} \citep{harris2020array}, a widely used package for numerical operations in Python. Parallelisation in \pysco is simplified: by replacing `range' with `prange' in loops, the code takes advantage of multi-core processing. \texttt{Numba} functions are typically compiled just-in-time (JIT), meaning they are compiled the first time the function is called. \texttt{Numba} infers input and output types dynamically, supporting function overloading for different types.

In \pysco, however, most functions are compiled ahead-of-time (AOT), meaning they are compiled as soon as the code is executed or imported. This is because the simulation uses 32-bit floating point precision for all fields, allowing for AOT compilation. Since the simulation operates on a uniform grid, unlike AMR simulations, there is no need for fine-grained grids. Therefore, using 32-bit precision is sufficient and does not result in any loss of accuracy. Additionally, 32-bit floats improve performance by enabling SIMD (Single Instruction, Multiple Data) instructions, which the compiler implicitly optimises for.

\subsection{Units and conventions}

We adopt the same strategy as \ramses and use supercomoving units \citep{martel1998convenient},  where the Poisson equation takes the same form as in classical Newtonian dynamics but includes a multiplicative scale factor
\begin{equation}
    \nabla^2\tilde{\phi} = \frac{3}{2}a\Omega_m\left(\tilde{\rho} - 1\right),
\end{equation}
where a tilde denotes a quantity in supercomoving units, $a$ is the scale factor, $\phi$ the gravitational potential, and $\rho$ the matter density.
We also define  conversion units from comoving coordinates and super-conformal time to physical SI units
\begin{align}
    \tilde{x} &= \frac{x}{x_*}, \quad \dd\tilde{t} = \frac{\dd t}{t_*}, \quad   \tilde{v} = v \frac{t_*}{x_*}, \\
    \tilde{\phi} &= \phi \frac{t_*^2}{x_*^2},  \quad \tilde{c} = c \frac{t_*}{ax_*}, 
\end{align}
where $\tilde{x}$, $\dd\tilde{t}$, $\tilde{v}$, $\tilde{\phi}$ and $\tilde{c}$ represent the particle position, time and velocity, gravitational potential and speed of light in simulation units. The conversion factors are defined as follows
\begin{align}
    x_* &= 100aL_{\rm box}/H_0 ,\\
    t_* &= a^2/H_0, \\
    \rho_*&= \Omega_m\rho_c/a^3,
\end{align}
as the length, time and density conversion units to km\footnote{While \ramses converts to cm.}, seconds and kg/m$^3$ respectively, where $L_{\rm box}$ is the box length in comoving coordinates and $H_0$ is the Hubble parameter today (in seconds). The particle mass is given by
\begin{equation}
    m_{\rm part} = \rho_*x_*^3/N_{\rm part},
\end{equation}
where $N_{\rm part}$ is the total number of particles in the simulation. 

\subsection{Data structure}
\label{sec:data_structure}

In this section, we discuss how \pysco handles the storage of particles and meshes using C-contiguous \texttt{NumPy} arrays. This approach was chosen for its simplicity and readability, allowing functions in \pysco to be easily reused in different contexts.

For particles, the position and velocity arrays are stored with the shape $[N_{\rm part}, 3]$, where the elements for each particle are contiguous in memory. This format is more efficient than using a shape of $[3, N_{\rm part}]$, particularly for mass assignment, where operations are performed particle by particle.
To further enhance performance, particles are ordered using Morton indices (also known as z-curve indices) rather than linear or random ordering. Morton ordering improves cache usage and, thus, increases performance. This ordering is applied every $N_{\rm reorder}$ steps, as defined by the user, to maintain good data locality and avoid performance losses (see also Appendix \ref{appendix:data_locality}).
We did not consider space-filling curves with better data locality properties (such as the Hilbert curve), because the associated encoding and decoding algorithms are much more computationally expensive, and Morton curves already provide excellent data locality. While more complex data structures are available (such as linked lists, fully-threaded trees, octrees or kdtrees), preliminary tests showed that using Morton-ordered NumPy arrays strikes a good balance between simplicity and performance, without the overhead of creating complex structures.

For scalar fields on the grid, arrays are stored with a shape $\left[N_{\rm cells}^{1/3}, N_{\rm cells}^{1/3}, N_{\rm cells}^{1/3}\right]$ using linear indexing. Although linear indexing does not offer optimal data locality, it is lightweight and does not require additional arrays to store indices. Moreover, it works well with predictable (optimizable) memory-access patterns, such as those used in stencil operators.
For vector fields, such as acceleration, the arrays have a shape $\left[N_{\rm cells}^{1/3}, N_{\rm cells}^{1/3}, N_{\rm cells}^{1/3}, 3\right]$, similar to the format used for particle arrays, to maintain consistency and performance.

\subsection{Initial conditions}

In this section, we describe how \pysco handles the generation of initial conditions for simulations, although it can also read from pre-existing snapshots in other formats. \pysco can load data directly from \ramses/\texttt{pFoF} format used in the RayGal simulations \citep{breton2019imprints, rasera2022raygal} or from \texttt{Gadget} format using the \texttt{Pylians} library \citep{villaescusa-navarro2018pylians}.
\pysco computes the time evolution of the scale factor, growth factors, and Hubble parameters, with the Astropy library \citep{astropy2022astropy} and internal routines (see Appendix \ref{appendix:growth_factors}).

To generate the initial conditions, the code requires a linear power spectrum $P(k, z = 0)$, which is rescaled by the growth factor at the initial redshift $z_{\rm ini}$. Additionally, \pysco generates a realisation of Gaussian white noise $W$, which is used to apply initial displacements to particles.

Some methods generate Gaussian white noise in configuration space, which is particularly useful for zoom simulations \citep{pen1997generating, sirko2005, bertschinger2001multiscale, prunet2008initial, hahn2013music}. However, in \pysco, the white noise is computed as
\begin{equation}
    W(\bm{k}) = A(\bm{k}) e^{i\theta(\bm{k})},
\end{equation}
with $A(\bm{k})$ an amplitude drawn from a Rayleigh distribution given by $\mathcal{R}_{\rm dist} = \sqrt{-\ln\mathcal{U}]0,1]}$ with $\mathcal{U}]0,1]$ a uniform random sampling between 0 and 1, and $\theta(\bm{k}) = \mathcal{U}]0,1]$. We also fix $W(\bm{k}) = \overline{W(-\bm{k})}$ where a bar denotes a complex conjugate, to ensure that the configuration-space field is real valued. In our case, since we consider a regular grid with periodic boundary conditions, we can generate the white noise directly in Fourier space.
An initial realisation of a density field is computed using 
\begin{equation}
\label{eq:delta_ini_fourier}
    \delta_{\rm ini}(\bm{k}) = (2\pi)^{3/2}\sqrt{P(k)} W(\bm{k}),
\end{equation}
with $k = |\bm{k}|$. This ensures  that we recover the Gaussian properties
\begin{align}
    &\ev{\delta_{\rm ini}(\bm{k})} = 0, \\
    &\ev{\delta_{\rm ini}(\bm{k}) \delta_{\rm ini}(\bm{k}')} = (2\pi)^3P(k)\delta_D(\bm{k} + \bm{k}'),
\end{align}
where $\delta_D$ is a Dirac delta and $\delta_{\rm ini}(\bm{0}) = 0$. 
We have also implemented the option to use `paired and fixed' initial conditions \citep{angulo2016cosmological}. 
This method greatly reduces cosmic variance by running paired simulations with opposite phases, at the cost of introducing some non-Gaussian features. The concept here is that instead of averaging the product of modes to match the power spectrum, the individual modes are set directly to $(2\pi)^{3/2}\sqrt{P(k)}$. 
In practice, the density field is used to compute the initial particle displacement from a homogeneous distribution, rather than directly sampling $\delta_{\rm ini}$. We use Lagrangian perturbation theory (LPT), with options for first-order 1LPT (also called `Zel’dovich approximation', \citealt{zeldovich1970gravitational}), second-order 2LPT \citep{scoccimarro1998transients, crocce2006transients}, or third-order 3LPT \citep{catelan1995lagrangian, rampf2012lagrangian}. The displacement field at the initial redshift $z_{\rm ini}$ up to third order is expressed as
\begin{equation}
    \Psi(z_{\rm ini}) = D_+^{(1)}\Psi^{(1)} + D_+^{(2)}\Psi^{(2)} + D_+^{(3a)}\Psi^{(3a)} +  D_+^{(3b)}\Psi^{(3b)} +  D_+^{(3c)}\Psi^{(3c)},
\end{equation}
with $D_+ \equiv D_+(z_{\rm ini})^{(1)}$ is the linear (first order) growth factor at the initial redshift and $\Psi^{(n)}$ are the different orders of the displacement field at $z = 0$, which can be written as \citep{michaux2021accurate}
\begin{align}
    \Psi^{(1)} &= -\nabla\phi^{(1)}, \\
    \Psi^{(2)} &= \nabla\phi^{(2)}, \\
    \Psi^{(3a)} &= \nabla\phi^{(3a)}, \\
    \Psi^{(3b)} &= \nabla\phi^{(3b)}, \\
    \Psi^{(3c)} &= \nabla\times\bm{A}^{(3c)},
\end{align}
with
\begin{align}
    \phi^{(1)} &= \nabla^{-2}\delta_{\rm ini}, \\
    \phi^{(2)} &= \frac{1}{2}\nabla^{-2}\left[\phi_{,ii}^{(1)}\phi_{,jj}^{(1)} - \phi_{,ij}^{(1)}\phi_{,ij}^{(1)}\right], \\
    \phi^{(3a)} &= \nabla^{-2}\left[\det\phi_{,ij}^{(1)}\right], \\
    \phi^{(3b)} &= \frac{1}{2}\nabla^{-2}\left[\phi_{,ii}^{(2)}\phi_{,jj}^{(1)} - \phi_{,ij}^{(2)}\phi_{ij}^{(1)}\right], \\
    \bm{A}^{(3c)} &= \nabla^{-2}\left[\nabla\phi_{,i}^{(2)} \times \nabla\phi_{,i}^{(1)}\right],
\end{align}
and
\begin{align}
    \nabla^{(2)}\phi^{(2)} = &\phi_{,xx}^{(1)} \left(\phi_{,yy}^{(1)}+\phi_{,zz}^{(1)}\right) + \phi_{,yy}^{(1)}\phi_{,zz}^{(1)} \nonumber\\ 
      &- \phi_{,xy}^{(1)}\phi_{,xy}^{(1)} - \phi_{,xz}^{(1)}\phi_{,xz}^{(1)} - \phi_{,yz}^{(1)}\phi_{,yz}^{(1)}, \\
    \nabla^{2}\phi^{(3a)} = &\phi_{,xx}^{(1)}\phi_{,yy}^{(1)}\phi_{,zz}^{(1)} + 2\phi_{,xy}^{(1)}\phi_{,xz}^{(1)}\phi_{,yz}^{(1)} \nonumber\\
     & -\phi_{,yz}^{(1)}\phi_{,yz}^{(1)}\phi_{,xx}^{(1)} - \phi_{,xz}^{(1)}\phi_{,xz}^{(1)}\phi_{,yy}^{(1)} - \phi_{,xy}^{(1)}\phi_{,xy}^{(1)}\phi_{,zz}^{(1)}, \\
     \nabla^2\phi^{(3b)} = &\frac{1}{2}\phi_{,xx}^{(1)}\left(\phi_{,yy}^{(2)}+\phi_{,zz}^{(2)}\right) + \frac{1}{2}\phi_{,yy}^{(1)}\left(\phi_{,xx}^{(2)}+\phi_{,zz}^{(2)}\right) \nonumber\\
     &+ \frac{1}{2}\phi_{,zz}^{(1)}\left(\phi_{,xx}^{(2)}+\phi_{,yy}^{(2)}\right) - \phi_{,xy}^{(1)}\phi_{,xy}^{(2)} - \phi_{,xz}^{(1)}\phi_{,xz}^{(2)} - \phi_{,yz}^{(1)}\phi_{,yz}^{(2)}, \\
     \nabla^2A_x^{(3c)} = &\phi_{,xy}^{(2)}\phi_{,xz}^{(1)} - \phi_{,xz}^{(2)}\phi_{,xy}^{(1)} \nonumber \\
     &+\phi_{,yz}^{(1)}\left(\phi_{,yy}^{(2)}-\phi_{,zz}^{(2)}\right) - \phi_{,yz}^{(2)}\left(\phi_{,yy}^{(1)}-\phi_{,zz}^{(1)}\right), \\
     \nabla^2A_y^{(3c)} = &\phi_{,yz}^{(2)}\phi_{,yx}^{(1)} - \phi_{,yx}^{(2)}\phi_{,yz}^{(1)} \nonumber \\
     &+\phi_{,xz}^{(1)}\left(\phi_{,zz}^{(2)}-\phi_{,xx}^{(2)}\right) - \phi_{,xz}^{(2)}\left(\phi_{,zz}^{(1)}-\phi_{,xx}^{(1)}\right), \\
     \nabla^2A_z^{(3c)} = &\phi_{,xz}^{(2)}\phi_{,yz}^{(1)} - \phi_{,yz}^{(2)}\phi_{,xz}^{(1)} \nonumber \\
     &+\phi_{,xy}^{(1)}\left(\phi_{,xx}^{(2)}-\phi_{,yy}^{(2)}\right) - \phi_{,xy}^{(2)}\left(\phi_{,xx}^{(1)}-\phi_{,yy}^{(1)}\right).
\end{align}
In practice, we compute all derivatives directly in Fourier space, since configuration-space finite-difference gradients would smooth the small scales and thus create inaccuracies in the initial power spectrum.
The second- and third-order contributions in the initial conditions can be prone to aliasing effects due to the quadratic and cubic non-linearities involved. To mitigate this, we apply Orszag’s 3/2 rule \citep{orszag1971elimination}, as suggested in \cite{michaux2021accurate}. The impact of this correction is minimal, by around 0.1\% on the power spectrum at small scales, when using 3LPT initial conditions and for a relatively late start with $z_{\rm ini}\approx 10$).
The initial position and velocity are then (up to third order)
\begin{align}
\label{eq:x_ini_LPT}
\bm{x}_{\rm ini} &= \bm{q} + \Psi^{(1)}D_+^{(1)} + \Psi^{(2)}D_+^{(2)} \\
 & + \Psi^{(3a)}D_+^{(3a)} + \Psi^{(3b)}D_+^{(3b)} + \Psi^{(3c)}D_+^{(3c)}, \\
\label{eq:v_ini_LPT}
    \bm{v}_{\rm ini} &= \Psi^{(1)}Hf^{(1)}D_+^{(1)} + \Psi^{(2)}Hf^{(2)}D_+^{(2)} \\ 
    & + \Psi^{(3a)}Hf^{(3a)}D_+^{(3a)} + \Psi^{(3b)}Hf^{(3b)}D_+^{(3b)} + \Psi^{(3c)}Hf^{(3c)}D_+^{(3c)},
\end{align}
with $H$ the Hubble parameter at $z_{\rm ini}$, $f_n$ the growth rate contribution to the $n$-th order, and $\bm{x}_{\rm uniform}$ the position of cell centres (or cell edges, see also Appendix~\ref{appendix:initial_positions}) when $N_{\rm part} = N_{\rm cells}$. We internally compute the growth factor and growth rate contributions as described in Appendix~\ref{appendix:growth_factors}. 

\subsection{Integrator}
\label{sec:integrator}

In our simulations, we employ the second-order symplectic Leapfrog scheme, often referred to as Kick-Drift-Kick, to integrate the equations of motion for the particles. The steps in the scheme are as follows
\begin{equation}
\begin{array}{llll}
\bm{v}_{i+1/2} & = & \bm{v}_{i} + \bm{a}_i \Delta t/2, & {\rm Kick}, \\
\bm{x}_{i+1} & = & \bm{x}_{i} + \bm{v}_{i+1/2} \Delta t, & {\rm Drift}, \\
\bm{v}_{i+1} & = & \bm{v}_{i+1/2} + \bm{a}_{i+1}\Delta t/2, & {\rm Kick}, \\
\end{array}
\end{equation}
where the subscript $i$ indicates the integration step, while $\bm{x}$, $\bm{v}$ and $\bm{a}$ are the particle positions, velocities and accelerations respectively. There are several ways to set the time step $\Delta t$. Some authors use linear or logarithmic spacing, depending on the user input. In our case, we followed a similar strategy as \ramses \citep{teyssier2002cosmological} and use several time stepping criteria. The first criterion is based on a cosmological time step that guarantees the scale factor does not change by more than a specified amount (by default 2\%, see also Appendix~\ref{appendix:time_stepping}), which is particularly effective at high redshift. The second criterion is based on the minimum free-fall time given by
\begin{equation}
    \Delta t_{\rm ff} = \sqrt{\frac{h}{{\rm max}(|\bm{a_i}|)}},
\end{equation}
with $h$ the cell size. We select the smallest value between this criterion and the cosmological time step. Additionally, we implemented a third criterion based on particle velocities $\Delta t_{\rm vel} = h/{\rm max}(|\bm{v_i}|)$, though we found this value often exceeds $\Delta t_{\rm ff}$ in practice. This approach ensures that the time step dynamically adapts to the structuration of dark matter in the simulation.
To further refine the time step, we multiply it by a user-defined Courant-like factor.

An interesting prospect for the future is the use of integrators based on Lagrangian perturbation theory (LPT) that could potentially reduce the number of time steps required while maintaining high accuracy (as suggested by \citealt{rampf2025bullfrog}). However, such methods couple the integration scheme with specific theories of gravity through growth factors. Since these factors may not always be accurately computed (for example, in MOND), we prefer to maintain the generality of the standard leapfrog integration scheme for now. We may explore the implementation of such LPT-based integrators in the future for specific theories of gravitation.

\subsection{Iterative solvers}

To displace the particles we first need to compute the force (or acceleration). There are various algorithms available for this purpose, either computing the force directly from a particle distribution, or determining the gravitational potential from which the force can subsequently be derived.
When using the gravitational potential approach, the force can be recovered by applying a finite-difference gradient operator $\bm{g} = -\nabla\phi$. Given that our results are sensitive to the order of the operator used, we have implemented several options for central difference methods, each characterised by specific coefficients. These coefficients are detailed in Table~\ref{tab:gradient_stencil}.
\begin{table}
  \caption{Stencil operator coefficients for central derivatives.}
    \smallskip
  \label{tab:gradient_stencil}
    \smallskip
  \begin{tabular}{|c|c|c|c|c|c|c|c|c|}
    \hline
    \rowcolor{blue!5}
    & & & & & & & & \\[-8pt]
    \rowcolor{blue!5}
    Points  & Accuracy & $-3$ & $-2$ & $-1$ & 0 & 1 & 2 & 3 \\
    \hline
        & & & & & & & & \\[-9pt]
    3 & $\mathcal{O}(h^2)$ & 0 & 0 & $-\frac{1}{2}$ & 0 & $\frac{1}{2}$ & 0 & 0 \\
        & & & & & & & & \\[-9pt]
    \hline
        & & & & & & & & \\[-9pt]
    5 & $\mathcal{O}(h^4)$ & 0 & $\frac{1}{12}$ & $-\frac{2}{3}$ & 0 & $\frac{2}{3}$ & $-\frac{1}{12}$ & 0 \\
        & & & & & & & & \\[-9pt]
    \hline
        & & & & & & & & \\[-9pt]
    7 & $\mathcal{O}(h^6)$ & $-\frac{1}{60}$ & $\frac{3}{20}$ & $-\frac{3}{4}$ & 0 & $\frac{3}{4}$ & $-\frac{3}{20}$ & $\frac{1}{60}$ \\
    \hline
  \end{tabular}
  \tablefoot{The columns refer to the stencil operator, accuracy and coefficients $c_n$ of the indices $n$ such as the gradient operator at the index $i$ can be written as $\nabla = \frac{1}{h}\sum_{n=-3}^{n=3} c_n u_{i+n}$, with $h$ the grid size.}
\end{table}
We aim to solve the following problem
\begin{equation}
\label{eq:abstract_elliptic_equation}
    \mathcal{L}u = f
\end{equation}
Where $u$ is unknown, $f$ is known and $\mathcal{L}$ is an operator. For the classical Poisson equation, $u \equiv \phi$, $f \equiv 4\pi G\rho$ and $\mathcal{L} \equiv \nabla^2$ with the seven-point Laplacian stencil
\begin{equation}
\label{eq:7pt-laplacian}
\nabla^2 u_{i,j,k} = \frac{1}{h^2}\left(L_{i,j,k}(u) - 6u_{i,j,k}\right),
\end{equation}
with the subscripts $i,j,k$ the cell indices and $L_{i,j,k}(u) = u_{i+1,j,k} + u_{i,j+1,k} + u_{i,j,k+1} + u_{i-1,j,k} + u_{i,j-1,k} + u_{i,j,k-1}$.

Lastly, $f$ is the density (source) term of Eq.~\eqref{eq:abstract_elliptic_equation}, which is directly estimated from the position of dark-matter particles. In code units, the sum of the density over the full grid must be 
\begin{equation}
    \sum_{i,j,k}\tilde{\rho}_{i,j,k} = N_{\rm part},
\end{equation}
where the density in a given cell is computed using the `nearest-grid point' (NGP), `cloud-in-cell' (CIC) or `triangular-shaped cloud' (TSC) mass-assignment schemes 
\begin{equation}
W\e{NGP}(x_i) = 
\left\lbrace
\begin{array}{lcl}
1  &  \rm{if} & |x_i| < 0.5,  \\
0& \rm{otherwise,} &  
\end{array}\right.
\end{equation}
\begin{equation}
W\e{CIC}(x_i) = 
\left\lbrace
\begin{array}{lcl}
1 - |x_i|  &  \rm{if} & |x_i| < 1.0,  \\
0& \rm{otherwise,} &  
\end{array}\right.
\end{equation}
\begin{equation}
W\e{TSC}(x_i) = 
\left\lbrace
\begin{array}{lcc}
0.75 - x_i^2  &  \rm{if} & |x_i| < 0.5,  \\
\left(1.5-|x_i|\right)^2/2 & \rm{else~if} & 0.5 < |x_i| < 1.5,  \\
0& \rm{otherwise.} &  
\end{array}\right.
\end{equation}
Here, $x_i$ is the normalised separation between a particle and cell positions, scaled by the cell size. In a three-dimensional space, this implies that a dark-matter particle contributes to the density of one, eight, or twenty-seven cells depending on whether the NGP, CIC, or TSC scheme is employed, respectively. For consistency, we use the same (inverse) scheme to interpolate the acceleration from cells to particles.

\subsection{The Jacobi and Gauss-Seidel methods}
\label{sec:jacobi_gs_methods}

Let us consider $\mathcal{L} = \begin{bmatrix}
    \ell_{11} & \ell_{12} & \ell_{13} & \dots  & \ell_{1n} \\
    \ell_{21} & \ell_{22} & \ell_{23} & \dots  & \ell_{2n} \\
    \vdots & \vdots & \vdots & \ddots & \vdots \\
    \ell_{n1} & \ell_{n2} & \ell_{n3} & \dots  & \ell_{nn}
\end{bmatrix}$, $u = \begin{bmatrix}
    u_1 \\
    u_2 \\
    \vdots \\
    u_n
\end{bmatrix}$ and $f = \begin{bmatrix}
    f_1 \\
    f_2 \\
    \vdots \\
    f_n
\end{bmatrix}$. The Jacobi method is a naive iterative solver which, for Eq.~\eqref{eq:abstract_elliptic_equation}, takes the form
\begin{align}
 \ell_{11} u_1^{\rm new} &+ \ell_{12} u_2^{\rm old} + \ell_{13} u_3^{\rm old} + \dots + \ell_{1n}u_n^{\rm old} = f_1, \\
 \ell_{21} u_1^{\rm old} &+ \ell_{22} u_2^{\rm new} + \ell_{23} u_3^{\rm old} + \dots + \ell_{2n}u_n^{\rm old} = f_2, \\
 \ell_{31} u_1^{\rm old} &+ \ell_{32} u_2^{\rm old} + \ell_{33} u_3^{\rm new} + \dots + \ell_{3n}u_n^{\rm old} = f_3, \\
 \vdots \nonumber\\
  \ell_{n1} u_1^{\rm old} &+ \ell_{n2} u_2^{\rm old} + \ell_{n3} u_3^{\rm old} + \dots + \ell_{nn}u_n^{\rm new} = f_n,
\end{align}
where the superscripts `new' and `old' refer to the iteration. This means that for one Jacobi sweep, the new value is directly given by that of the previous iteration. 

In practice, we use the Gauss-Seidel method instead, which has a better convergence rate and lower memory usage. This method is akin to Jacobi’s but enhances convergence by incorporating the most recently updated values in the computation of the remainder. It follows
\begin{align}
 \ell_{11} u_1^{\rm new} &+ \ell_{12} u_2^{\rm old} + \ell_{13} u_3^{\rm old} + \dots + \ell_{1n}u_n^{\rm old} = f_1, \\
 \ell_{21} u_1^{\rm new} &+ \ell_{22} u_2^{\rm new} + \ell_{23} u_3^{\rm old} + \dots + \ell_{2n}u_n^{\rm old} = f_2, \\
 \ell_{31} u_1^{\rm new} &+ \ell_{32} u_2^{\rm new} + \ell_{33} u_3^{\rm new} + \dots + \ell_{3n}u_n^{\rm old} = f_3, \\
 \vdots \nonumber \\
  \ell_{n1} u_1^{\rm new} &+ \ell_{n2} u_2^{\rm new} + \ell_{n3} u_3^{\rm new} + \dots + \ell_{nn}u_n^{\rm new} = f_n,
\end{align}
which seems impossible to parallelise because  each subsequent equation relies on the results of the previous one. However, we implement a strategy known as `red-black' ordering in the Gauss-Seidel method. This technique involves colouring the cells like a chessboard, where each cell is designated as either red or black. When updating the red cells, we only utilise information from the black cells, and vice versa. This approach is equivalent to performing two successive Jacobi sweeps—one for the red cells and another for the black cells.
For the Laplacian operator as expressed in Eq.~\eqref{eq:7pt-laplacian}, the Jacobi sweep for a given cell can be formulated as follows
\begin{equation}
\label{eq:jacobi_sweep_laplacian}
    u_{i,j,k}^{\rm new} = \frac{1}{6}\left(L_{i,j,k}\left(u^{\rm old}\right) - h^2f_{i,j,k}\right).
\end{equation}
In scenarios involving a non-linear operator, finding an exact solution may not be feasible. In such cases, we linearise the operator using the Newton-Raphson method, expressed as
\begin{equation}
 \label{eq:newtonraphson}
    u^{\rm new} = u^{\rm old} - \frac{\mathcal{L}(u^{\rm old})}{\partial\mathcal{L}/\partial u^{\rm old}},
\end{equation}
This approach is commonly referred to as the `Newton Gauss-Seidel method', allowing for more effective convergence in non-linear contexts.

\subsection{Multigrid}
\label{sec:multigrid}
In practice, both the Jacobi and Gauss-Seidel methods are known for their slow convergence rates, often requiring hundreds to thousands of iterations to reach a solution and typically unable to achieve high accuracy. To overcome these limitations, a popular and efficient iterative method known as `multigrid' is employed. This algorithm significantly accelerates convergence by solving the equation iteratively on coarser meshes, effectively addressing large-scale modes.
The multigrid algorithm follows this procedure \citep{press1992numericalrecipes}: we first discretise the problem on a regular mesh with grid size $h$ as
\begin{equation}
    \mathcal{L}_h u_h = f_h,
\end{equation}
which can be solved using the Gauss-Seidel method, and where $\mathcal{L}_h$ is the numerical operator on the mesh with grid size $h$ which approximates $\mathcal{L}$. $\tilde{u}_h$ denotes our approximate solution and $v_h$ is the `error' on the true solution
\begin{equation}
    v_h = u_h - \tilde{u}_h,
\end{equation}
and the `residual' is given by
\begin{equation}
    d_h = f_h - \mathcal{L}\tilde{u}_h.
\end{equation}
Depending on whether the operator $\mathcal{L}$ is linear or non-linear, different multigrid schemes will be employed to solve the equations effectively.

\subsubsection{Linear multigrid}

Considering the case where $\mathcal{L}$ is linear, meaning that $\mathcal{L}_h(u_h - \tilde{u}_h) = \mathcal{L}_hu_h - \mathcal{L}_h\tilde{u_h}$, we have the relation
\begin{equation}
\label{eq:operator_error_residual_h}
    \mathcal{L}_hv_h = d_h.
\end{equation}
From there, the goal is to estimate $v_h$ to find $u_h$. We use numerical methods to find the approximate solution $\tilde{v}_h$ by solving
\begin{equation}
\label{eq:multigrid_error_equation}
    \mathcal{L}_h\tilde{v}_h = d_h,
\end{equation}
using Gauss-Seidel. The updated approximation for the field is 
\begin{equation}
    \tilde{u}_h^{\rm new} = \tilde{u}_h + \tilde{v}_h.
\end{equation}
The issue is that the approximate operator $\mathcal{L}_h$ is usually local and finite-difference based, for which long-range perturbations are slowly propagating and therefore very inefficient computationally. The multigrid solution to this issue is to solve the error on coarser meshes to speed up the propagation of long-range modes. 
First, we use the `restriction' operator $\mathcal{R}$ which interpolates from fine to coarse grid
\begin{equation}
    d_H = \mathcal{R}d_h,
\end{equation}
where $H = 2h$ is the grid size of the coarse mesh. We then solve Eq.~\eqref{eq:multigrid_error_equation} on the coarser grid to infer $\tilde{v}_H$ as 
\begin{equation}
    \label{eq:coarser_operator_error}
        \mathcal{L}_H\tilde{v}_H = d_H.
\end{equation}
We then use the prolongation operator $\mathcal{P}$, which interpolates from coarse to fine grid
\begin{equation}
    \tilde{v}_h = \mathcal{P}\tilde{v}_H,
\end{equation}
and we finally update our approximation on the solution
\begin{equation}
    \tilde{u}_h^{\rm new} = \tilde{u}_h + \tilde{v}_h.
\end{equation}
We have provided a brief overview of the multigrid algorithm using two grids, but in practice, to solve for $\tilde{v}_H$, we can extend the scheme to even coarser meshes. This results in a recursive algorithm where the coarser level in \pysco contains $16^3$ cells. There are several strategies for navigating the different mesh levels, commonly referred to as V, F, or W-cycles. The V cycle is the simplest and quickest to execute, although it converges at a slower rate than the F and W cycles (see Appendix~\ref{appendix:multigrid_convergence}).

For the restriction and prolongation operators, the lowest-level schemes employ averaging (where the field value on the parent cell is the mean of its children's values) and straight injection (where child cells inherit the value of their parent), respectively. However, as noted by \cite{guillet2011simple}, to minimise inaccuracies in the estimation of the final solution, we use a higher-order prolongation scheme defined as
\begin{equation}
\mathcal{P} = 
\left\lbrace
\begin{array}{ll}
27/64, & x_{0,1,2} < 0.5, \\
9/64, & x_{0,1} < 0.5 < x_2 < 1.5, \\
3/64, & x_0 < 0.5 < x_{1,2} < 1.5,   \\
1/64, & 0.5 < x_{0,1,2} < 1.5, \\
0, & {\rm otherwise}, \\
\end{array}\right.
\end{equation}
where $\bm{x} = (x_0, x_1, x_2)$ with $x_0 \leq x_1 \leq x_2$, is the separation (normalised by the fine grid size) between the centre of the fine and coarse cells. 

We consider our multigrid scheme to have converged when the residual is significantly lower than the `truncation error', defined as:
\begin{equation}
    \tau \equiv \mathcal{L}_h u  - f_h,
\end{equation}
and which can be estimated as \citep{press1992numericalrecipes, li2012ecosmog}
\begin{equation}
\tilde{\tau}_h \equiv \mathcal{L}_H(\mathcal{R}\tilde{u}_h) - \mathcal{R}\mathcal{L}_h\tilde{u}_h.
\end{equation}
We consider that we reached convergence when 
\begin{equation}
\label{eq:multigrid_threshold}
    |d_h| < \alpha \tilde{\tau}_h,
\end{equation}
where $|d_h|$ is the square root of the quadratic sum over the residual in each cell of the mesh, $\alpha$ is the stopping criterion. It is noteworthy that \citep{knebe2001MLAPM} proposed using an alternative approach for estimating the truncation error
\begin{equation}
    \tau_{h, K01} = \mathcal{P}\left(\mathcal{L}_H(\mathcal{R}\tilde{u}_h)\right) - \mathcal{L}_h\tilde{u}_h,
\end{equation}
which we can approximate by 
\begin{equation}
   \tau_{h, K01} \approx \mathcal{P}\left(\mathcal{R}f_h\right) - f_h, 
\end{equation}
to reduce computational time, when the source term is non-zero. The relation between these two approaches is $\tau_h \approx 0.1\tau_{h, K03}$. In practice, we use the first estimation $\tau_h$.
Since the Jacobi and Gauss-Seidel methods are iterative, an initial guess is still required. While the multigrid algorithm is generally insensitive to the initial guess in most cases, some small optimisations can be made. For instance, if we initialise the full grid with zeros, one Jacobi step directly provides
\begin{equation}
\label{eq:multigrid_initial_guess}
    u^{\rm ini}_{i,j,k} = -\frac{h^2}{6} f_{i,j,k},
\end{equation}
which is used to initialise $u_h$ for the first simulation step, and $v_H$ the error on coarser meshes. 

In addition to the Gauss-Seidel method, \ramses also incorporates `successive over-relaxation' (SOR), which allows the updated field to include a contribution from the previous iteration, expressed as $\phi^{n+1} = \omega\phi^{n+1} + (1-\omega)\phi^{n}$, 
where $n$ denotes the iteration step, and $\omega$ is the relaxation factor.
Typically, we perform two Gauss-Seidel sweeps before the restriction (termed pre-smoothing) and one sweep after the prolongation (post-smoothing), controlled by the parameters \texttt{Npre} and \texttt{Npost}. We set $\omega = 1.25$ by default (see also Appendix~\ref{appendix:multigrid_convergence}), similarly to \cite{kravtsov1997ART}.

\begin{figure}
\centering
\includegraphics[width=\hsize]{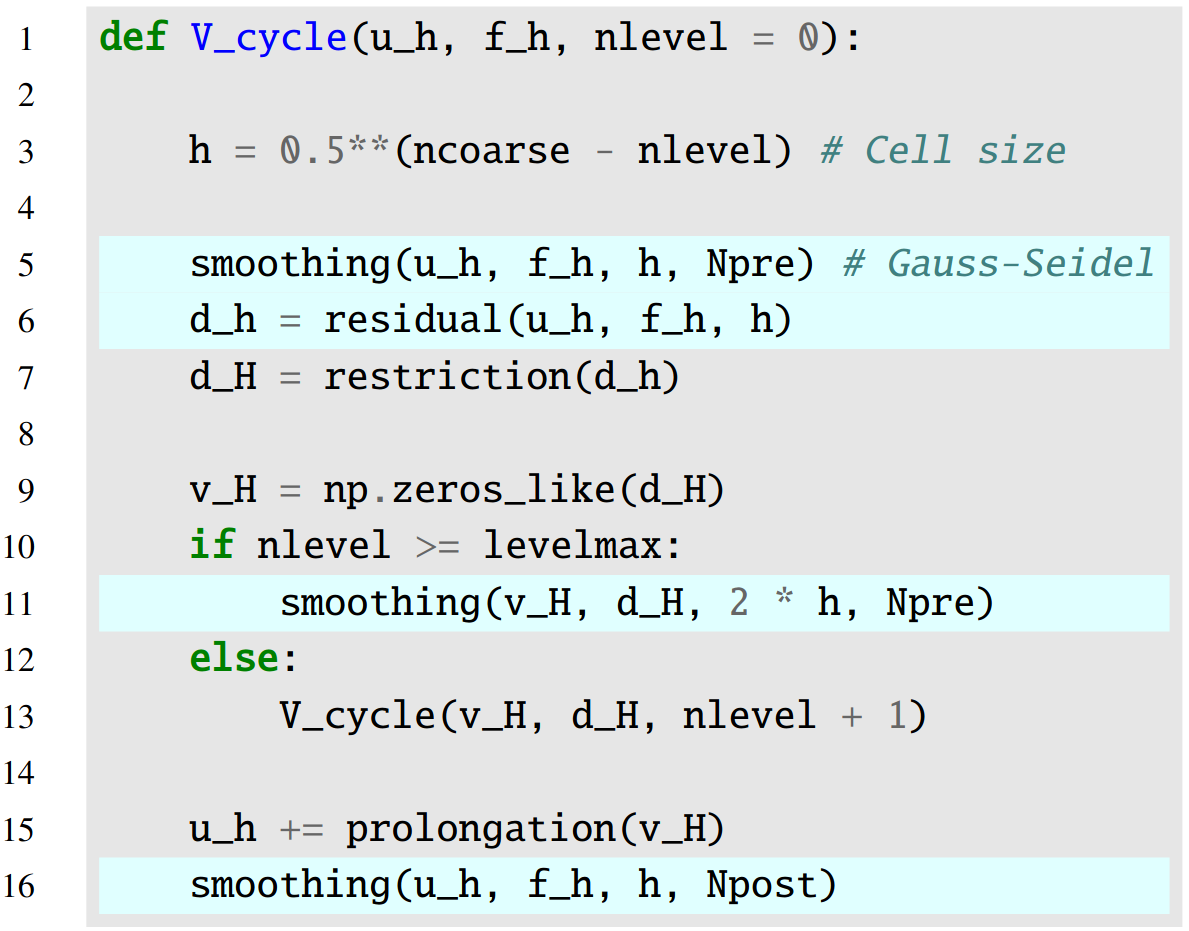}
\caption{Pseudo code in Python for the multigrid V-cycle algorithm. We highlight in cyan the lines where the theory of gravitation enters.}
\label{code:V_cycle}
\end{figure}

A pseudo code for the V cycle is depicted in Fig.~\ref{code:V_cycle},
where the components that require modification based on the gravitational theory are the \texttt{smoothing} and \texttt{residual} functions. 
As noted by \cite{press1992numericalrecipes}, omitting SOR (setting $\omega = 1$) for the first and last iterations can enhance performance since this allows the prolongation operator (line 15) to operate on only half the mesh, since the other half will be directly updated during the first Gauss-Seidel sweep. Similarly, the restriction operator can also be applied to only half of the mesh because, following one Gauss-Seidel sweep, the residual will be zero on the other half by design.
It is thus feasible to combine lines 6 and 7 (the residual and restriction operators) to compute the coarser residual from half of the finer mesh.
In this example, we use a zero initial guess for the error at the coarser level, but in principle, we could initialise it with a Jacobi step (as shown in Eq.~\ref{eq:multigrid_initial_guess} for Newtonian gravity).

This optimisation, however, should be approached cautiously, as the residual is zero up to floating-point precision. In certain scenarios, this can significantly impact the final solution. For example, in \pysco, we primarily utilise 32-bit floats, rendering this optimisation less accurate for the largest modes. Consequently, we ultimately decided to retain only the optimisation concerning the prolongation on half the mesh when employing the non-linear multigrid algorithm, as we do not utilise SOR in this context (see Appendix~\ref{appendix:multigrid_convergence}).

\subsubsection{Non-linear multigrid}
\label{sec:non-lineary_multigrid}
If $\mathcal{L}$ is a non-linear operator instead, then we need to solve for
\begin{equation}
   \mathcal{L}_h(\tilde{u}_h + v_h) - \mathcal{L}_h\tilde{u}_h = d_h.
\end{equation}
Contrarily to the linear case in Eq.~\eqref{eq:coarser_operator_error} where we only need to solve for the error at the coarser level, we now use
\begin{equation}
    \mathcal{L}_H \tilde{u}_H = \mathcal{L}_H \left(\mathcal{R}\tilde{u}_h\right) + \mathcal{R} d_h,
\end{equation}
where we need to store the full approximation of the solution $\tilde{u}$ at every level, hence the name `Full Approximation Storage'. Finally, we update the solution as 
\begin{equation}
    \tilde{u}^{\rm new}_h = \tilde{u}_h + \mathcal{P}\left(\tilde{u}_H - \mathcal{R}\tilde{u}_h\right).
\end{equation}

\subsection{Fast Fourier transforms}
\label{sec:fft_methods}

We implemented three different fast Fourier transforms (FFT) procedures to compute the force field, most of which were already implemented in \texttt{FastPM} \citep{feng2016fastpm}.

\textbf{- FFT}: The Laplacian operator is computed through the Green's function kernel \\
\begin{equation}
    \nabla^{-2} = -k^{-2} W_{\rm MAS}^{-2}(k),
\end{equation}
where $W_{\rm MAS}(k)$ is the mass-assignment scheme filter, given by
\begin{equation}
    W_{\rm MAS}(k) = \left[\prod_{d = x,y,z} \textrm{sinc}\left(\frac{\omega_d}{2}\right)\right]^p,
\end{equation}
with $\omega_d = k_d h$ between $[-\pi, \pi]$ and where $p = 1, 2, 3$ for NGP, CIC and TSC respectively \citep{hockney1981computer, jing2005correcting}.

 \textbf{- FFT\_7pt}: Instead of using the exact kernel for the Laplacian operator, we use the Fourier-space equivalent of the seven-point stencil Laplacian (Eq.~\ref{eq:7pt-laplacian}), which reads
\begin{equation}
    \nabla^{-2} = -\left[\sum_{d=x,y,z} \left(h\omega_d\textrm{sinc}\frac{\omega_d}{2}\right)^2\right]^{-1},
\end{equation}
where no mass-assignment compensation is used. For both FFT and FFT\_7pt methods, the force is estimated through the finite-difference stencil as shown in Table~\ref{tab:gradient_stencil}.

\textbf{- FULL\_FFT}: The force is directly estimated in Fourier space through the differentiation kernel
\begin{equation}
    \nabla\nabla^{-2} = -i\bm{k}k^{-2} W_{\rm MAS}^{-2}(k).
    \label{eq:exact_fullfft_kernel}
\end{equation}
As we will see in section~\ref{sec:comparison_to_ramses}, this naive operator can become very inaccurate when the simulation has more cells than particles.

\subsection{Newtonian and parametrised simulations}
\label{sec:newtonian_simulations}
In Eq.~\eqref{eq:multigrid_initial_guess}, we showed how to provide a generic initial guess for the multigrid algorithm. However, leveraging our understanding of the underlying physics can enable us to formulate an even more accurate first guess, thus reducing the number of multigrid cycles required for convergence.
In the context of $N$-body simulations, we anticipate that the potential field will closely resemble that of the preceding step, especially when using sufficiently small time steps. This allows us to adopt the potential from the previous step as our initial guess for the multigrid algorithm. While this approach facilitates faster convergence to the true solution, it necessitates storing one additional grid in memory.
Moreover, it is important to note that, in the linear regime in Newtonian cosmology, the density contrast evolves as a function of redshift with a scale-independent growth factor. Consequently, we can optimise our first guess by rescaling the potential field from the previous step according to the following equation
\begin{equation}
\label{eq:rescaming_newtonian_potential}
    \tilde{\phi}(z_1) = \frac{(1+z_1) D_+(z_1)}{(1+z_0) D_+(z_0)} \tilde{\phi}(z_0),
\end{equation}
where $z_0$ and $z_1$ denote the initial and subsequent redshifts, respectively. This rescaling is performed for every time step of the simulation, except for the initial time step, ensuring that we maintain an efficient and accurate estimate for our initial guess in the multigrid process. More details can be found in Appendix.~\ref{appendix:multigrid_convergence}.

\subsection{MOND simulations}

\begin{figure}
\centering
\includegraphics[width=1.0\hsize]{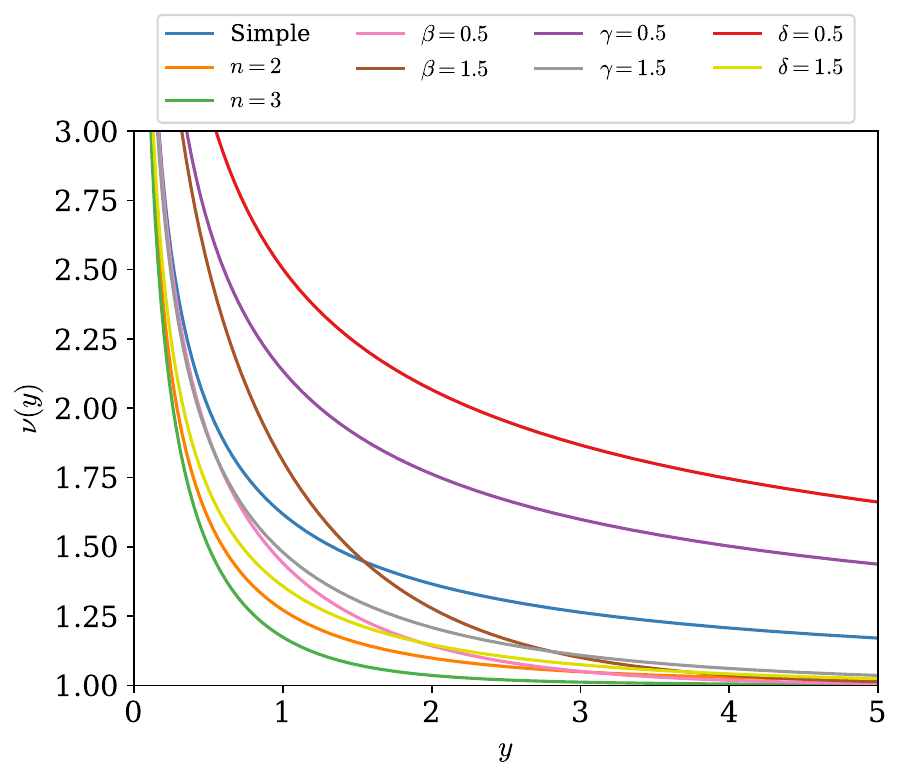}
\caption{MOND interpolating functions for the different families and parameters shown in Sec.~\ref{sec:theory_mond}.}
\label{fig:mond_interpolating_functions}
\end{figure}

The classical formulations of MOND (such as AQUAL and QUMOND) have already been implemented in several codes \citep{nusser2002mond,llinares2008cosmological, angus2012qumond,candlish2015raymond,lughausen2015por, visser2024fast}. In \pysco, we further allow, if specified by the user, for a time-dependent acceleration scale $g_0 \rightarrow a^\mathcal{N}g_0$, which basically delays (or accelerates) the entry of perturbations in the MOND regime, and which is set to $\mathcal{N} = 0$ by default. In Fig.~\ref{fig:mond_interpolating_functions}, we show the behaviour of the interpolating functions $\nu(y)$ described in Sec.~\ref{sec:theory_mond}.  
We see that, as expected from Eq.~\eqref{eq:QUMOND_Poisson_MOND}, $\nu(y) \rightarrow 1$ when $y \gg 1$, meaning that we recover Newtonian gravity in a regime of strong acceleration.

To implement the right-hand side of Eq.~\eqref{eq:QUMOND_Poisson_MOND}, 
 we employ a method analogous to that described by \cite{lughausen2015por}, and using the same notation
\begin{align}
\label{eq:QUMOND_Poisson_MOND_discrete}
        \nabla^2\phi^{\rm N} = \frac{1}{h}&\Big[\nu_{B_x}\nabla\left(\phi^{\rm N}\right)_{B_x,x} - \nu_{A_x}\nabla\left(\phi^{\rm N}\right)_{A_x,x} \nonumber\\
    &+ \nu_{B_y}\nabla\left(\phi^{\rm N}\right)_{B_y,y} - \nu_{A_y}\nabla\left(\phi^{\rm N}\right)_{A_y,y} \nonumber\\
    &+ \nu_{B_z}\nabla\left(\phi^{\rm N}\right)_{B_z,z} - \nu_{A_z}\nabla\left(\phi^{\rm N}\right)_{A_z,z}\Big],
\end{align}
where $B_i$ and $A_i$ are the points at $+0.5h ~\bm{e}_i$ and $-0.5h~\bm{e}_i$ respectively, with $\bm{e}_i \in \{\bm{e}_x, \bm{e}_y, \bm{e}_z\}$, the unit vectors of the simulation. We have also defined
\begin{equation}
    \nu_{B_x} = \nu\left(\frac{\sqrt{\left[\nabla\left(\phi^{\rm N}\right)_{B_x,x}\right]^2 + \left[\nabla\left(\phi^{\rm N}\right)_{B_x,y}\right]^2 + \left[\nabla\left(\phi^{\rm N}\right)_{B_x,z}\right]^2}}{g_0}\right),
\end{equation}
with $\nabla\left(\phi^{\rm N}\right)_{B_x,i}$ the $i$-th component of the force (with a minus sign) at the position $B_x$. The force components are estimated as
\begin{align}
\label{eq:}
    \nabla(\phi)_{B_x,x} &= \frac{\phi^{\rm N}_{1,0,0} - \phi^{\rm N}_{0,0,0}}{h}, \\
    \nabla(\phi)_{B_x,y} &= 
    \frac{\left(\phi^{\rm N}_{1,1,0} - \phi^{\rm N}_{1,-1,0} \right) + \left(\phi^{\rm N}_{0,1,0} - \phi^{\rm N}_{0,-1,0} \right)}{4h} \\
    \nabla(\phi)_{B_x,z} &= 
    \frac{\left(\phi^{\rm N}_{1,0,1} - \phi^{\rm N}_{1,0,-1} \right) + \left(\phi^{\rm N}_{0,0,1} - \phi^{\rm N}_{0,0,-1} \right)}{4h},   
\end{align}
and similarly for other points. 
We note that for $B_x,x$ (as well as for $B_y,y$ and $B_z,z$), we perform a three-point central derivative with half the mesh size, differing from the approach taken by \citep{lughausen2015por}, who implemented a non-uniform five-point stencil. This decision was made to ensure that in the Newtonian case (that is, $\nu = 1$), we exactly recover the seven-point Laplacian, maintaining consistency with the Laplacian operator employed in our multigrid scheme. Consequently, we opted to retain three-point derivatives for the other components as well (although we can still use a different stencil order when computing the acceleration from the MOND potential).

Additionally, we exclusively solve Eq.~\eqref{eq:QUMOND_Poisson_MOND_discrete} using either the multigrid or FFT\_7pt solvers. Using the FFT solver presents challenges because, although we deconvolve $\phi^{\rm N}$ by the mass-assignment scheme kernel, the uncorrected mesh discreteness in the force computation introduces inaccuracies that can significantly affect the matter power spectrum.

\subsection{$f(R)$ simulations}

In supercomoving units, the $f(R)$ field equations from Eqs.~\eqref{eq:fr_theory_1}-\eqref{eq:fr_theory_2} are given by \citep{li2012ecosmog}
\begin{align}
    \label{eq:fr_sim_1}
    &\nabla^2\tilde{\phi} = \frac{3}{2} \Omega_m a \left(\tilde{\rho} - 1\right) - \frac{\tilde{c}^2}{2}\nabla^2\tilde{f}_R, \\
    \label{eq:fr_sim_2}    
    &\nabla^2\tilde{f}_R = -\frac{1}{\tilde{c}^2}\Omega_m a \left(\tilde{\rho} - 1\right) + \frac{1}{3\tilde{c}^2}\bar{R}a^4\left[ \left(\frac{\bar{f}_R}{\tilde{f}_R}\right)^{1/(n+1)} - 1\right],
\end{align}
where $\tilde{f}_R = a^2f_R\tilde{c}^2/c^2$. While Eq.~\eqref{eq:fr_sim_1} is linear and can be solved by standard techniques, Eq.~\eqref{eq:fr_sim_2} is not and needs some special attention. To this end, \cite{oyaizu2008nonlinear} used a non-linear multigrid algorithm (see Section~\ref{sec:non-lineary_multigrid}) with the Newton-Raphson method (as shown in Eq.~\ref{eq:newtonraphson}), making the change of variable $u \equiv \ln\left(\tilde{f_R}/\bar{f}_R\right)$ to avoid unphysical zero-crossing of $f_R$.
\cite{bose2017speeding} proposed instead  that for this specific model, one could perform a more appropriate change of variable, that we generalise here to $u \equiv \left(\tilde{f}_R/\bar{f}_R\right)^{1/(n+1)}$. We can then recast Eq.~\eqref{eq:fr_sim_2} as
\begin{equation}
\label{eq:fr_depressed_equation}
    u^{n+1} + pu + q = 0,
\end{equation}
where 
\begin{align}
    p &= \frac{h^2}{6\tilde{c}^2\bar{f}_R}\left[\Omega_m a\left(1 - \tilde{\rho}\right) - \frac{a^4\bar{R}}{3}\right] - \frac{1}{6}L_{i,j,k}\left(u^{n+1}\right), \\
    q &= \frac{a^4h^2\bar{R}}{18\tilde{c}^2\bar{f}_R}.
\end{align}
We note that $q$ is necessarily negative (because $\bar{f}_R < 0$), which is useful to determine the branch of the solution for Eq.~\eqref{eq:fr_depressed_equation}.

\textbf{- Case $n = 1$:} as noticed in \cite{bose2017speeding}, when making the change of variable $u = \sqrt{-f_{R}}$, the field equation could be recast as a depressed cubic equation,
\begin{equation}
    u^3 + pu + q = 0\,,
    \label{eq:cubic_equation}
\end{equation}
which possesses the analytical solutions  \citep{ruan2022fast}
\begin{equation}
u =  
\left\lbrace
\begin{array}{ll}
\left(-q\right)^{1/3}, & p = 0, \\
-\left[C + \Delta_0/C\right]/3, & p > 0, \\
-\left[C + \Delta_0/C\right]/3, & p < 0~\rm{ and }~ \Delta_1^2 - 4\Delta_0^3 > 0, \\
 -2\sqrt{\Delta_0} \cos\left(\theta/3 + 2\pi/3\right)/3, & \rm{else}, 
\end{array}\right.
\end{equation}
with $\Delta_0 = -3p$, $\Delta_1 = 27q$, $C = \left[\frac{1}{2}\left(\Delta_1 + \sqrt{\Delta_1^2 - 4\Delta_0^3}\right)  \right]^{1/3}$ and $\cos\theta = \Delta_1/\left(2\Delta_0^{3/2}\right)$.

\textbf{- Case $n = 2$:} the field equation can be rewritten as a quartic equation \citep{ruan2022fast}
\begin{equation}
    u^4 + pu + q = 0,
\end{equation}
with the roots
\begin{equation}
u =  
\left\lbrace
\begin{array}{ll}
-S + \frac{1}{2}\sqrt{-4S^2 + p/S}, & p > 0, \\
\left(-q\right)^{1/4}, & p = 0, \\
S + \frac{1}{2}\sqrt{-4S^2 - p/S}, & p < 0, \\
\end{array}\right.
\end{equation}
where $S = \frac{1}{2}\sqrt{\frac{1}{3}\left(Q + \Delta_0/Q\right)}$, $Q = \left(\frac{1}{2}\left[\Delta_1 + \sqrt{\Delta_1^2 - 4\Delta_0^3}\right]\right)^{1/3}$, $\Delta_1 = 27p^2$ and $\Delta_0 = 12q$. 
Due to non-zero residuals in our multigrid scheme, the $q$ term can become positive.
This situation results in inequalities such as $\Delta_1^2 - 4\Delta_0^3 < 0$, which lack an analytical solution, or $Q + \Delta_0/Q < 0$. In both cases, we enforce $u = (-q)^{1/4}$.

While the Gauss-Seidel smoothing procedure remains necessary (as $p$ depends on the values of the field $u$ in adjacent cells), this method eliminates the requirement for the Newton-Raphson step and the computationally expensive exponential and logarithmic operations employed in the \citep{oyaizu2008nonlinear} method, resulting in significant performance enhancements. Given that the operations needed to determine the branch of solutions for cubic and quartic equations are highly sensitive to machine precision, we conduct all calculations using 64-bit floating-point precision.

We must use the non-linear multigrid algorithm outlined above to solve the scalaron field, with its initial guess provided directly by the solution from the previous step, without any rescaling since we are solving for $u$ rather than $f_R$. Because the tolerance threshold is heavily dependent on redshift, we cannot apply the same criterion used for the linear Poisson equation; by default, we consider convergence to be achieved after one F cycle.
In fact, we do not solve Eq. (113) directly; instead, we incorporate the $f(R)$ contribution during force computation as follows
\begin{equation}
    \bm{F} = \bm{F}_{\rm Newton} + \frac{\tilde{c}^2}{2}\nabla\tilde{f}_R.
\end{equation}
This choice was made because replacing $\nabla^2\tilde{f}_R$ in Eq.~\eqref{eq:fr_sim_1} with Eq.~\eqref{eq:fr_sim_2} could lead to a right-hand side that has a non-zero mean due to numerical inaccuracies, resulting in artificially large residuals that cannot be reduced below the error threshold.

We direct interested readers to \cite{winther2015modified, adamek2024KPJC6P1}, along with references therein, for comparisons of numerical methods used to solve the modified Poisson equation in \cite{hu2007models} $f(R)$ gravity.

\section{Results}
\label{sec:results}

We consider a $\Lambda$CDM linear power spectrum computed by \texttt{CAMB} \citep{lewis2000efficient} with parameters $h = 0.7$, $\Omega_m = 0.3$, $\Omega_b = 0.05$, $\Omega_r = 8.5\cdot10^{-5}$, $n_s = 0.96$ and $\sigma_8 = 0.8$.
We run simulations with $512^3$ particles and as many cells (unless specifically stated) within a box of $256~h^{-1}$Mpc. It leads to a spatial resolution of $0.5~h^{-1}$Mpc (which is then reduced to $0.25~h^{-1}$Mpc and $0.125~h^{-1}$Mpc when using 1024$^3$ and 2048$^3$ cells, respectively). All the power spectrum results are shown for the snapshot at $z = 0$, and computed with a simple (without dealiasing) estimator implemented within \pysco.
Point-mass tests are also shown in Appendix~\ref{appendix:point_mass_test}.

\subsection{Initial conditions}
\label{sec:results_initial_conditions}
Ideally, the statistical properties of $N$-body simulations at late times should be independent of initial conditions, but studies have shown this is not the case. For example, \cite{crocce2006transients} suggested that using 2LPT instead of 1LPT could allow simulations to begin at a later time, reducing computational effort. Similar approaches were extended to 3LPT by \citep{michaux2021accurate} and to fourth-order (4LPT) by \citep{list2024starting}, which detailed improvements in particle resampling and aliasing mitigation.

This section systematically examines how the perturbative order, starting redshift, Poisson solver, and gradient order for force computations affect the results. Particles are initialised at cell centres, with results for initialisation at cell edges shown in Appendix~\ref{appendix:initial_positions}.

In Fig.~\ref{fig:initial_conditions_gradients},  the ratio of the matter power spectrum at $z = 0$ for various starting redshifts is compared to a reference simulation starting at $z_{\rm ini} = 150$. 
\begin{figure}
\centering
\includegraphics[width=1.0\hsize]{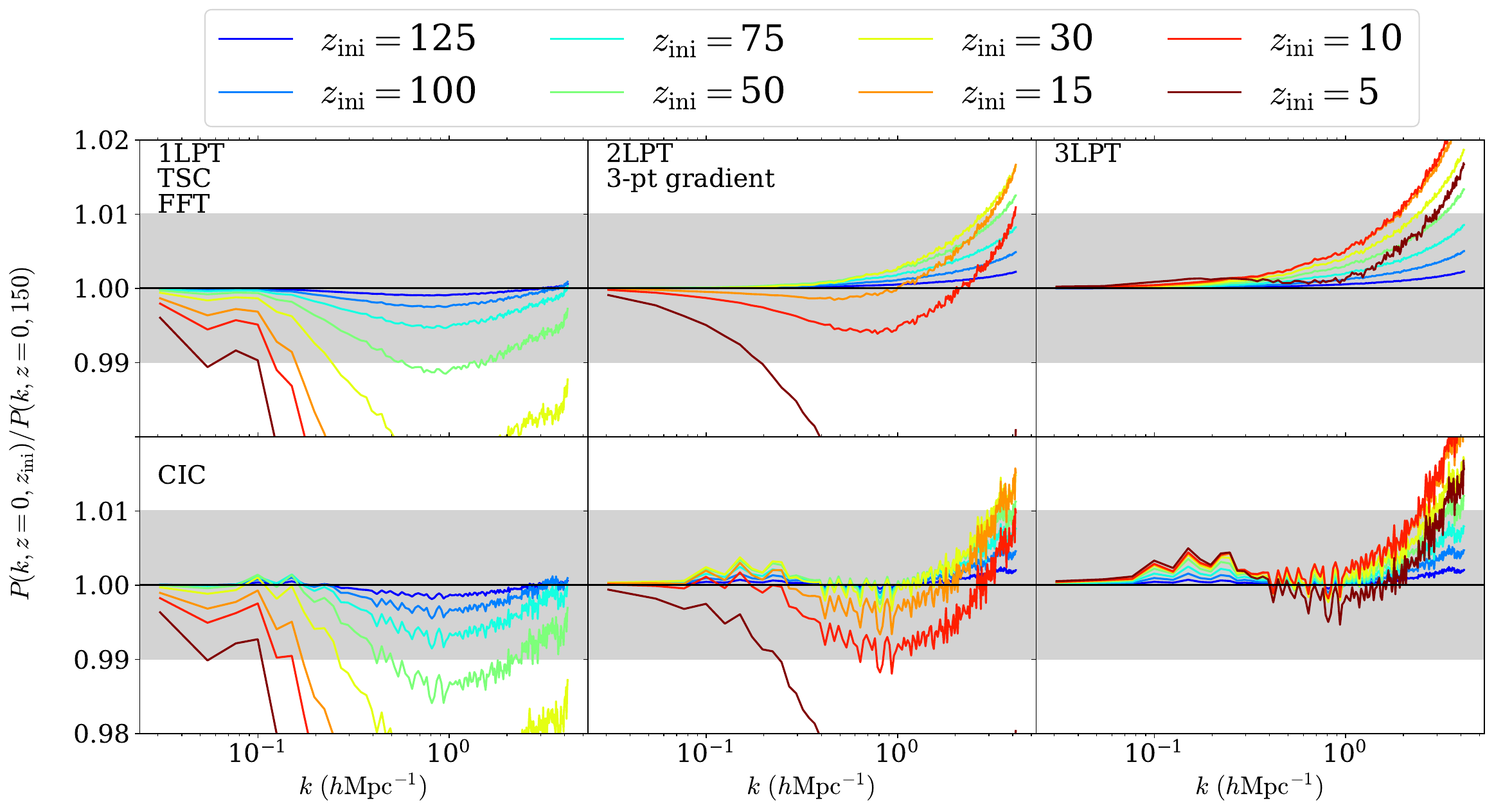}
\includegraphics[width=1.0\hsize]{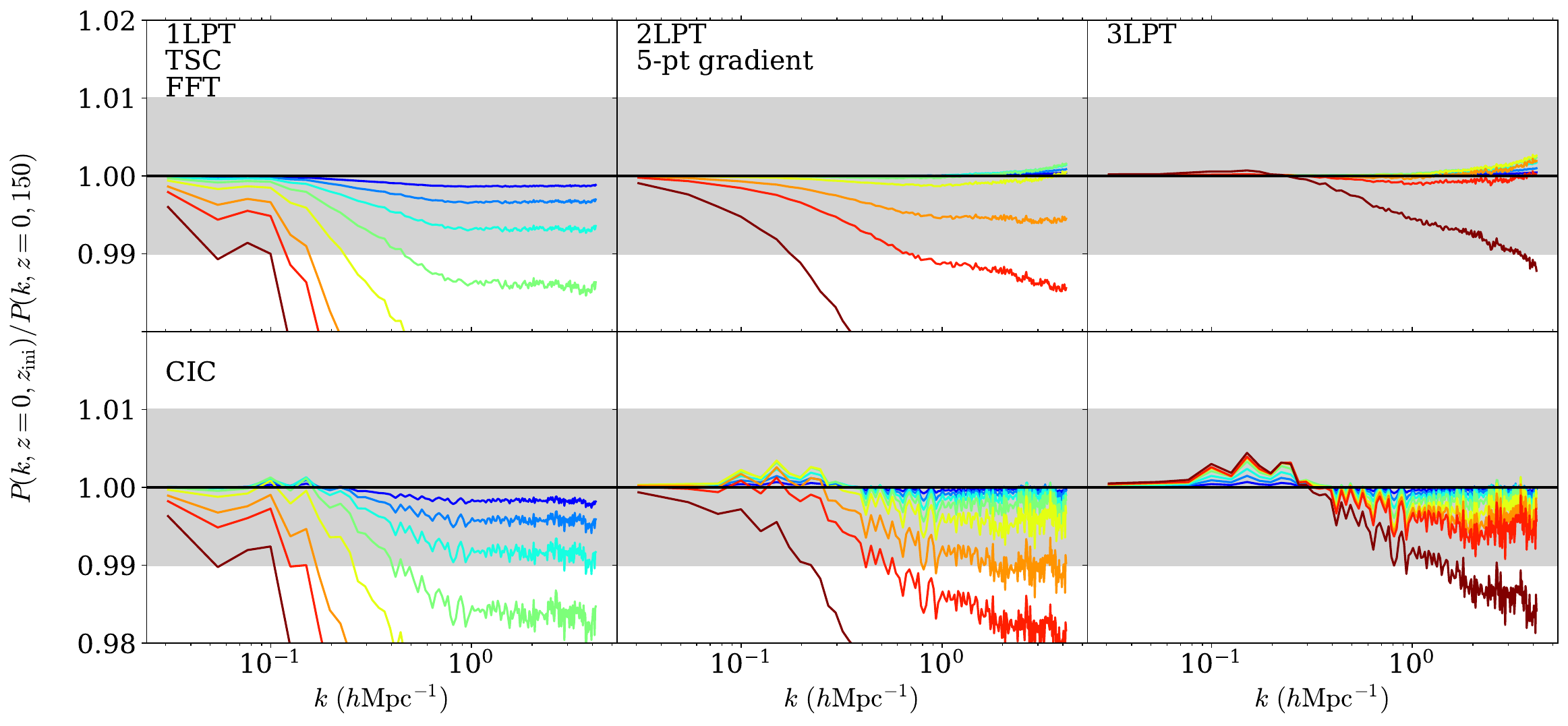}
\includegraphics[width=1.0\hsize]{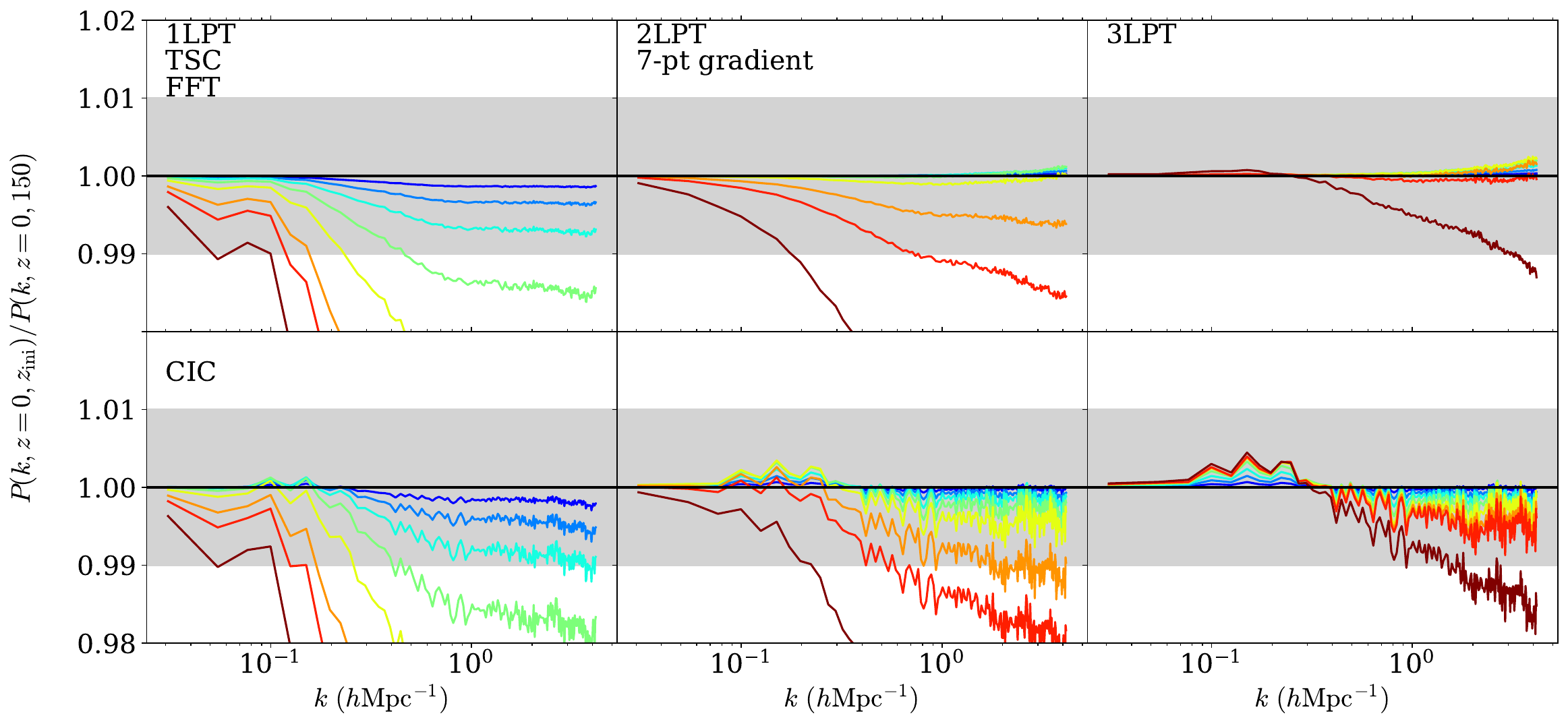}
\caption{Ratio of the matter power spectrum at $z = 0$ for different starting redshift $z_{\rm ini}$ (in coloured lines), with respect to a reference simulation with $z_{\rm ini} = 150$. We use 1LPT, 2LPT or 3LPT initial conditions from left to right and three-, five- and seven-point gradients from top to bottom panels. We use the FFT solver in any case (see Section~\ref{sec:fft_methods}). In each subplot the top and bottom panels show simulations which use TSC and CIC mass-assignments schemes respectively. Grey shaded area denotes a 1\% discrepancy w.r.t the reference power spectrum.}
\label{fig:initial_conditions_gradients}
\end{figure}
It indicates that using five- and seven-point gradient methods produces nearly identical results, while the three-point gradient shows significant deviations for both 2LPT and 3LPT. Thus, at least a five-point gradient is necessary to achieve convergence regarding the influence of initial conditions on late-time clustering statistics. Additionally, the CIC interpolation method results in more scattered data compared to the TSC method, which produces smoother density fields and is less prone to large variations in potential and force calculations. The results show a remarkable agreement within 0.1\% at $z_{\rm ini} \geq 125, 30$ and 10 for 1LPT, 2LPT and 3LPT respectively. This contrasts with \cite{michaux2021accurate}, where power suppression at intermediate scales was observed before increasing around the Nyquist frequency. In the present study, a maximum wavenumber $k_{\rm max} = 2k_{\rm Nyq}/3$ was used to avoid aliasing, yielding excellent agreement even at $z_{\rm ini} = 10$ with 3LPT. However, small-scale agreement can break down if these scales are not well-resolved, for instance, if the simulation box size increases but the number of particles and cells is kept constant.

\begin{figure}
\centering
\includegraphics[width=1.0\hsize]{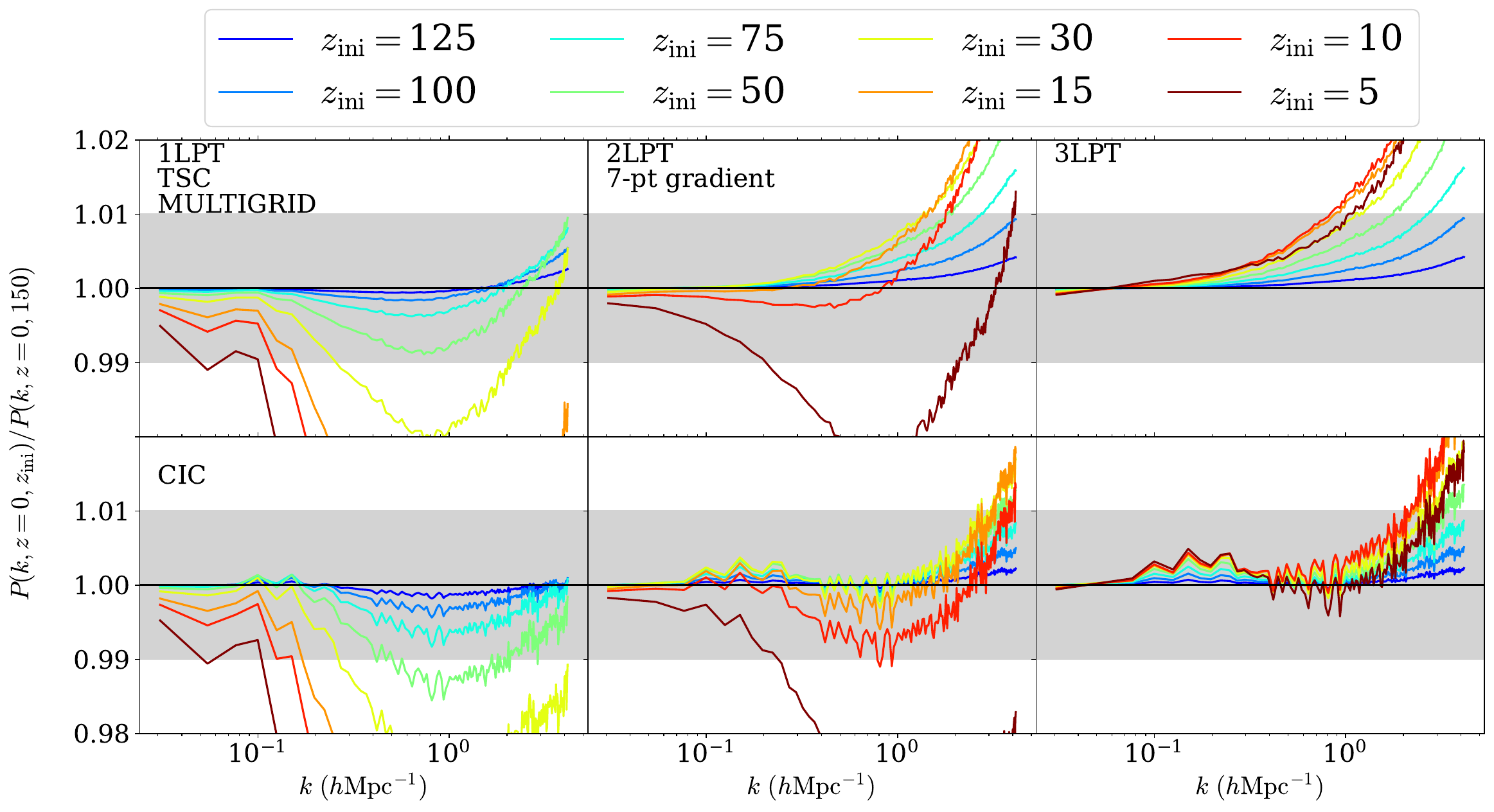}
\includegraphics[width=1.0\hsize]{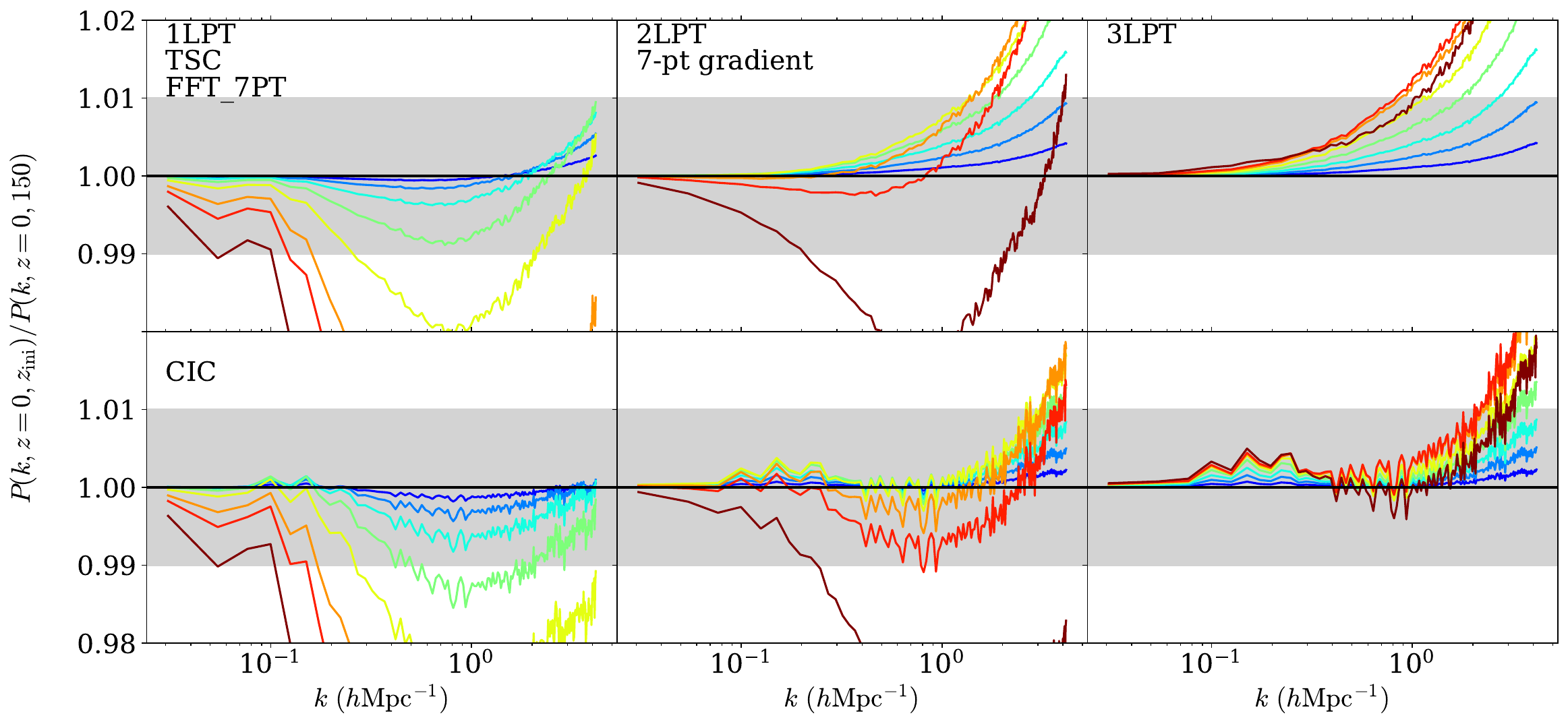}
\includegraphics[width=1.0\hsize]{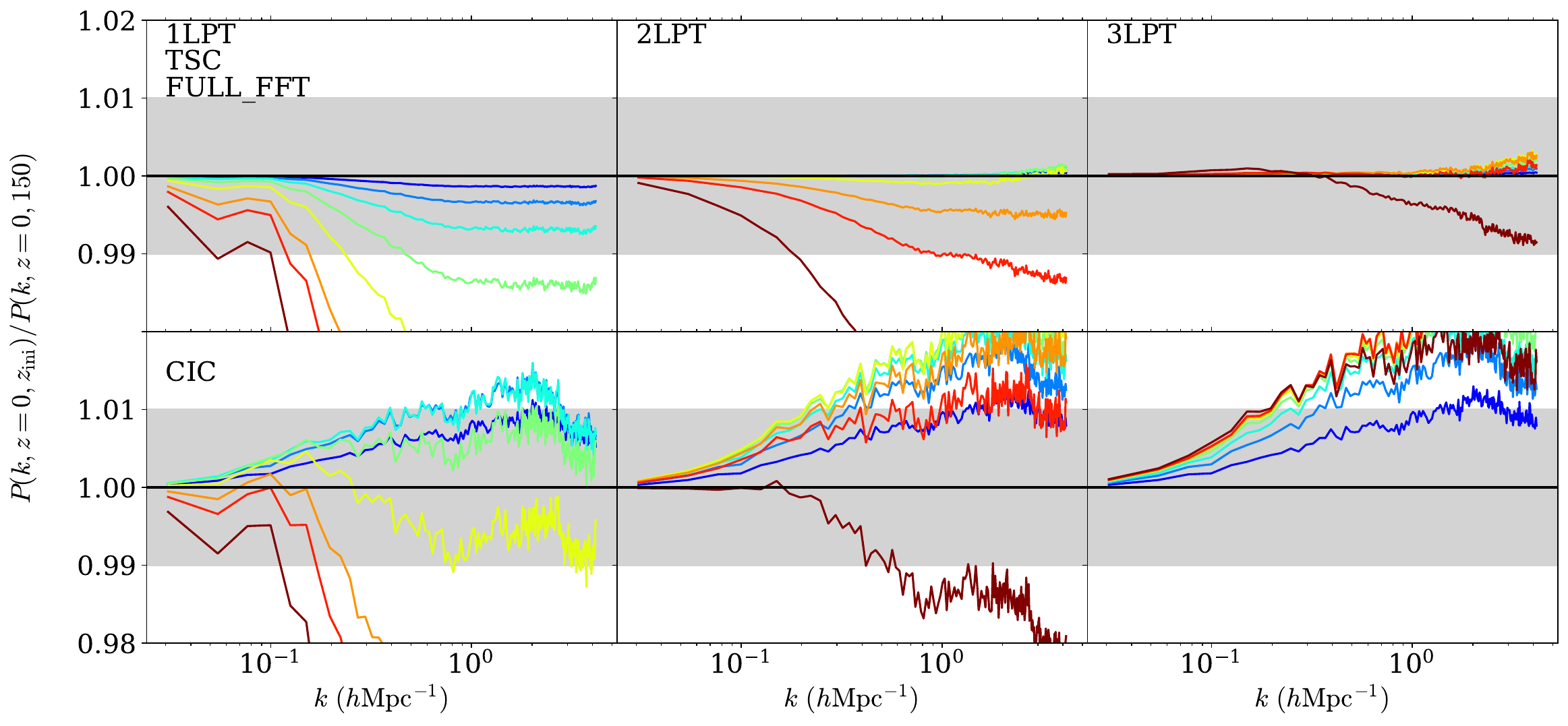}
\caption{Same as Fig.~\ref{fig:initial_conditions_gradients}, but varying the $N$-body solver instead of gradient order, which is kept to seven-point (except for the FULL\_FFT solver which does not make use of finite gradients). }
\label{fig:initial_conditions_solvers}
\end{figure}

Fig.~\ref{fig:initial_conditions_solvers} explores variations in the $N$-body solver. 
Multigrid and FFT\_7pt solvers produce nearly identical results, as both assume a seven-point Lagrangian operator, though Multigrid computes it in configuration space, while FFT\_7pt does so in Fourier space. Simulations initialised at later times exhibit excess small-scale power compared to the FFT solver in Fig.~\ref{fig:initial_conditions_gradients}, likely due to their poorer resolution of small scales compared to FFT (similarly to the effect of the three-point gradient stencil, as we will see in Section~\ref{sec:comparison_to_ramses}). The FULL\_FFT solver shows similar results to FFT when a TSC scheme is applied. However, using CIC with FULL\_FFT fails to achieve convergence, likely due to the solver’s sensitivity to the smoothness of the density field, as discussed in Section~\ref{sec:comparison_to_ramses}.

These findings suggest that simulations must use at least a five-point gradient to yield accurate results, with the ideal starting redshift depending on the LPT order. Additionally, employing TSC provides smoother results, and the FULL\_FFT solver should be avoided with the CIC scheme. Finally, to achieve good convergence at all scales, simulations need to adequately resolve the small scales.
This thus validates our implementation of LPT initial conditions in \pysco.

\subsection{Comparison to \ramses}
\label{sec:comparison_to_ramses}

This section compares \pysco with \ramses by running a PM-only \ramses simulation (disabling AMR) starting at $z_{\rm ini} = 50$ with 2LPT initial conditions generated by \texttt{MPGRAFIC} \citep{prunet2008initial}. The same initial conditions are used for all comparisons between \pysco and \ramses.

\begin{figure}
\centering
\includegraphics[width=1.0\hsize]{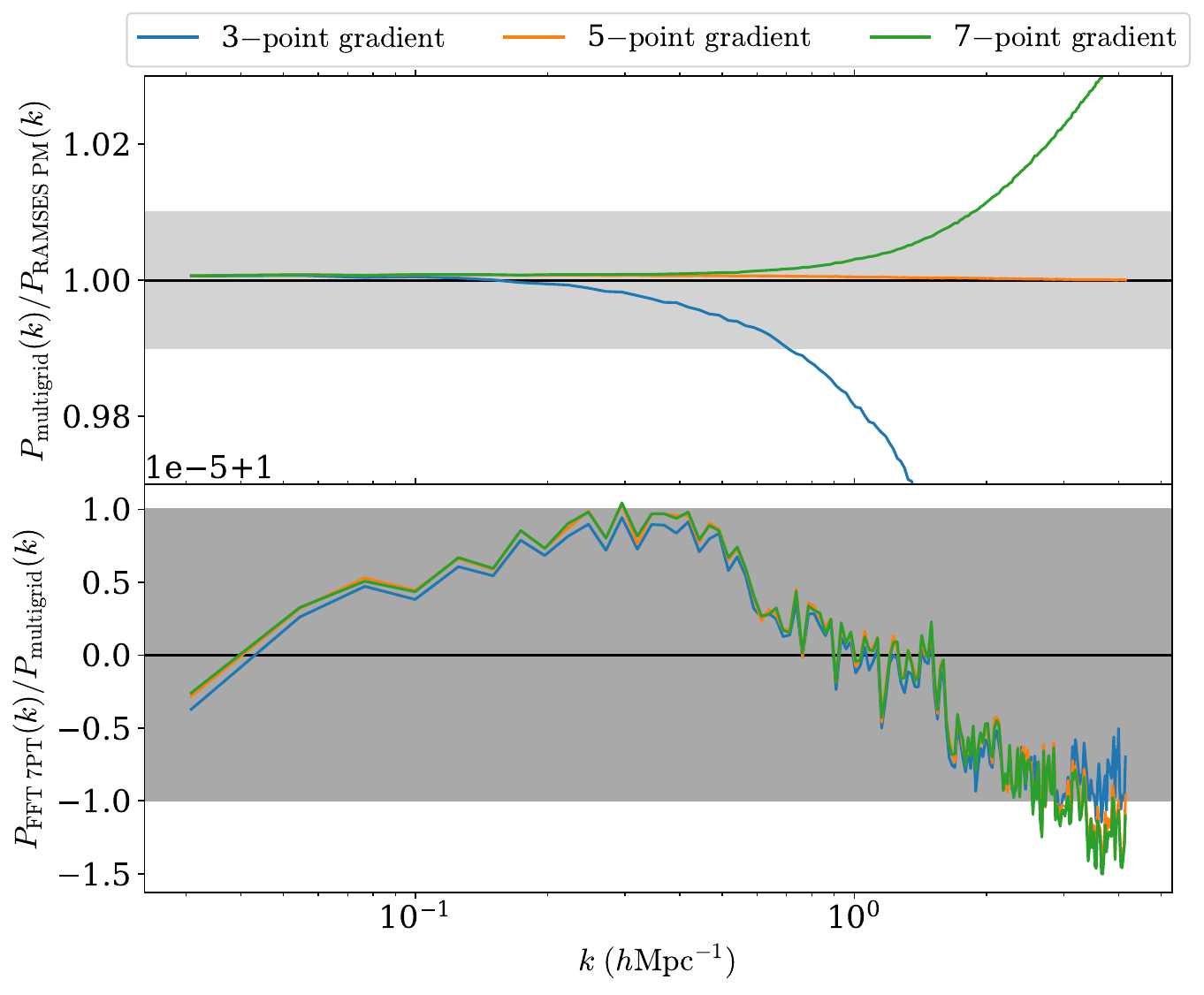}
\caption{Top panel: Ratio of the power spectrum for three-, five- and seven-point gradient in blue, orange and green lines respectively using multigrid, with respect to \ramses (with AMR disabled and also using multigrid). Because \ramses does not stop exactly at $z = 0$ we rescaled its power spectrum the linear growth factor. The grey shaded area show the $\pm 1\%$ limits. Bottom panel: Ratio of the power spectrum using the \pysco FFT\_7pt solver w.r.t \pysco multigrid. The dark grey shaded area show the $10^{-5}$ limits. In any case we use a TSC algorithm.}
\label{fig:pk_solvers_ramses_pm}
\end{figure}

In Fig.~\ref{fig:pk_solvers_ramses_pm}, we observe remarkable agreement between \pysco with a five-point gradient and \ramses (which also uses a five-point stencil with multigrid), with differences at only the 0.01\% level. This validates the multigrid implementation in \pysco.
Using a three-point gradient leads to a significant damping of the power spectrum at small scales, while the seven-point gradient shows an increase at even smaller scales. Based on these results and those from Section~\ref{sec:results_initial_conditions}, it is clear that a three-point stencil is suboptimal, as the small runtime gain is outweighed by the power loss at small scales.
Also shown in Fig.~\ref{fig:pk_solvers_ramses_pm} is the power spectrum ratio between the multigrid and FFT\_7pt solvers, with both solvers agreeing at the $10^{-5}$ level independently of the gradient order. This confirms the agreement already seen in Fig.~\ref{fig:initial_conditions_solvers}. Small fluctuations around unity could be due to the convergence threshold of the multigrid algorithm (set at $\alpha = 0.005$). Given this close agreement, FFT\_7pt results are not shown further in this paper, except for performance analysis in Section~\ref{sec:results_performances}.

\begin{figure}
\centering
\includegraphics[width=1.0\hsize]{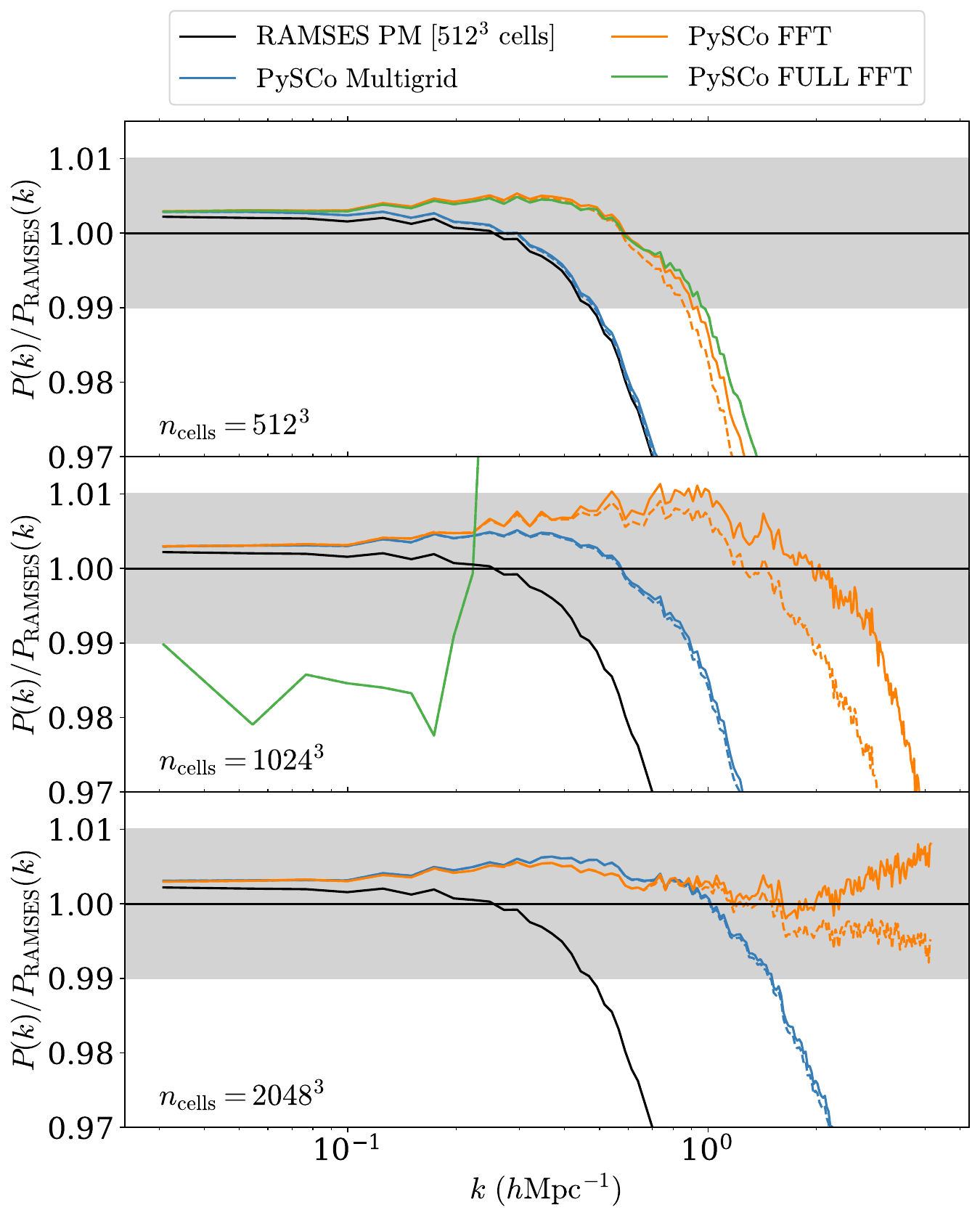}
\caption{Ratio of the power spectrum with respect to \ramses (with AMR). In black lines we show results for \ramses PM (no AMR), while in blue, orange and green we show results for \pysco using multigrid, FFT and FULL\_FFT solvers respectively. In top, middle and bottom panels we use $512^3$, $1024^3$ and $2048^3$ cells respectively for \pysco. In dashed and solid lines we plot results using five- and seven-point gradient operator. The grey shaded area show the $\pm 1\%$ limits. We use TSC in any case, and we do not plot FULL\_FFT for $n_{\rm cells} = 2048^3$.}
\label{fig:pk_solvers_ramses_amr_tsc}.
\end{figure}

\begin{figure}
\centering
\includegraphics[width=1.0\hsize]{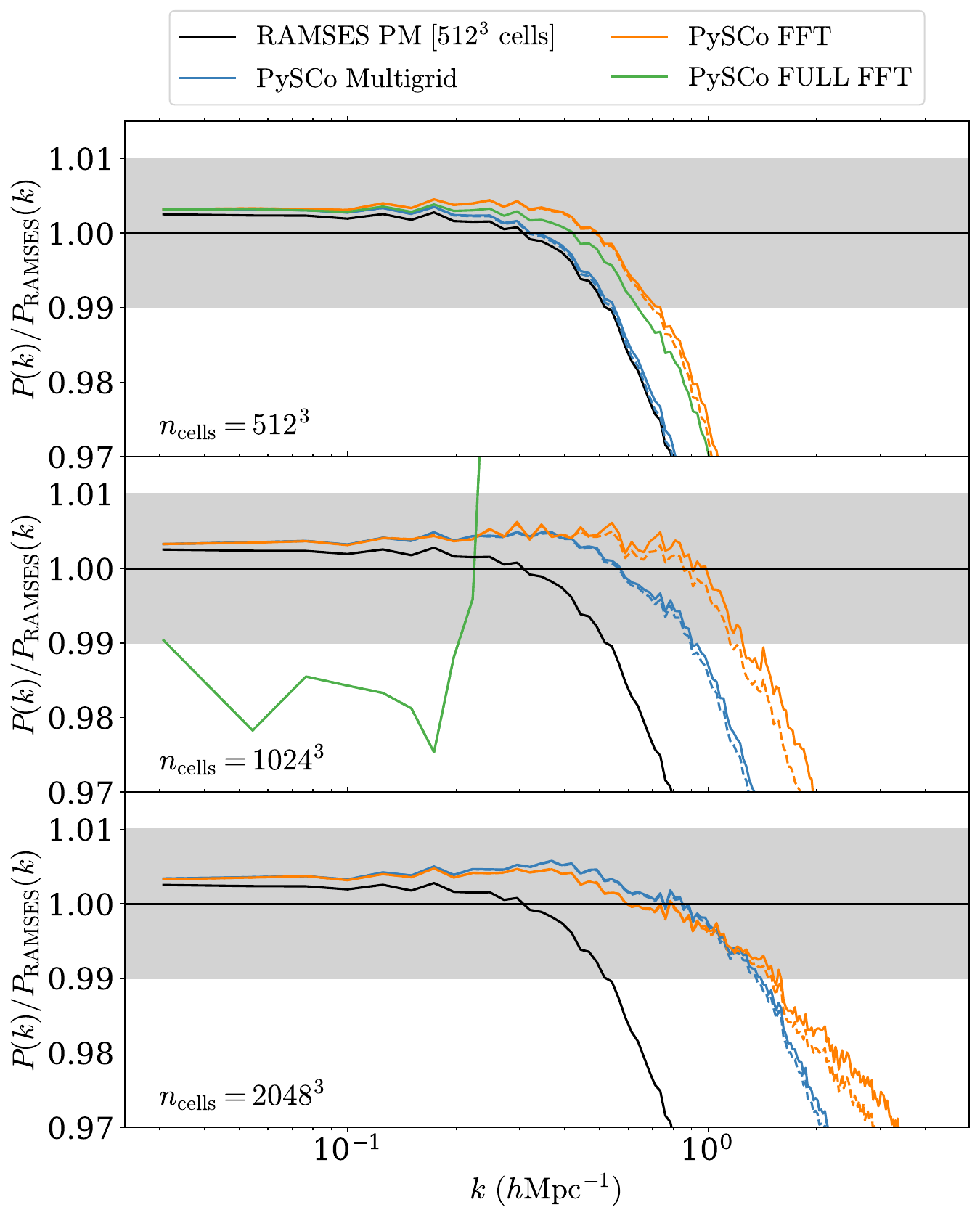}
\caption{Same as Fig.~\ref{fig:pk_solvers_ramses_amr_tsc} but using CIC.}
\label{fig:pk_solvers_ramses_amr_cic}
\end{figure}

In Fig.~\ref{fig:pk_solvers_ramses_amr_tsc}, the comparison between \ramses (AMR) and \pysco reveals that using a seven-point Laplacian operator in a PM-only code results in a significant suppression of small-scale power compared to AMR (as seen in \ramses PM, as well as \pysco multigrid and FFT\_7pt).
 However, using FFT or FULL\_FFT solvers improves small-scale resolution by a factor of two in wavenumbers before resolution effects become significant. With $n_{\rm cells} = 1024^3$, the FULL\_FFT solver fails entirely due to large scatter in the density grid, which contains eight times more cells than particles. Therefore, FULL\_FFT can only be used with a smooth field, where $n_{\rm part} \geq n_{\rm cells}$ and when a TSC scheme is employed. For other cases, a more sophisticated approach would be required, such as computing the mass-assignment kernel in configuration space for force computation and then Fourier-transforming it \citep{hockney1981computer}, which would be computationally expensive.
For $n_{\rm cells} = 1024^3$ and $2048^3$, \pysco with multigrid gains factors of 2 and 4 in wavenumbers, respectively, as expected. The approximated seven-point Laplacian operator already smooths the field significantly, so there is little difference between using five- or seven-point gradients. Using the FFT solver instead leads to more accurate small scales. For $n_{\rm cells} = 1024^3$, using a seven-point gradient achieves higher wavenumbers than a five-point gradient. With $n_{\rm cells} = 2048^3$, FFT agrees with RAMSES at the percent level across the full range, although the plots are restricted to $k_{\rm max} = 2k_{\rm Nyq}/3$ (where $k_{\rm Nyq}$ assumes $n_{\rm cells} = 512^3$).
In Fig.~\ref{fig:pk_solvers_ramses_amr_cic}, similar results are presented using CIC instead of TSC (also for the reference \ramses simulation).
The conclusions for multigrid and FULL\_FFT remain unchanged. However, the wavenumber gain with FFT is smaller compared to TSC, particularly for $n_{\rm cells} = 2048^3$, where multigrid and FFT exhibit similar behaviour and deviate from \ramses at the same scale. This indicates that for FFT, a smooth field is critical, though to a lesser extent than for FULL\_FFT.

To achieve the most accurate results compared to \ramses, it is necessary to use the FFT solver with TSC and a seven-point stencil for the gradient operator. Otherwise, a five-point gradient can be used without a loss of accuracy. We also remark a slight shift on large scales between the reference (AMR case) and PM runs. Because the initial conditions are the same and the cosmological tables very similar, we expect this small difference (roughly 0.1\%) to come from the fact that for RAMSES we do not stop exactly at $z = 0$ (although we correct analytically for that), and that the time stepping is also slightly different, as for the AMR case we enter in a regime where the free-fall time step dominates over the cosmological time step at earlier times (see also Appendix~\ref{appendix:time_stepping}).

\subsection{$f(R)$ gravity}
\begin{figure}
\centering
\includegraphics[width=1.0\hsize]{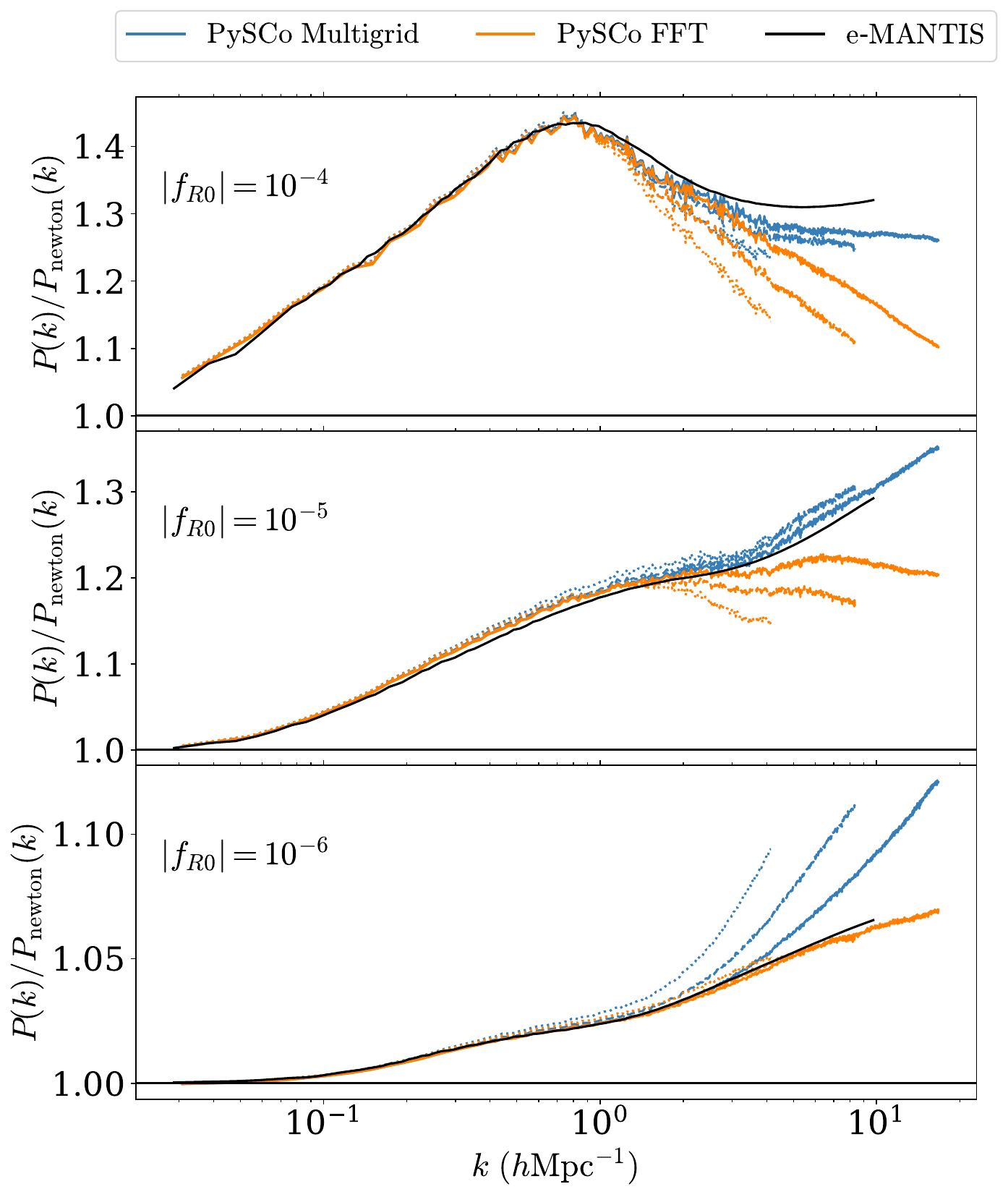}
\caption{Power spectrum boost from $f(R)$ gravity w.r.t to the Newtonian case. Blue and orange lines refer respectively to multigrid and FFT Poisson solvers with \pysco (in any case, the scalaron field equation is solved with non-linear multigrid), while black lines show the \emantis emulator. In dotted, dashed and solid lines we use $n_{\rm cells} = 512^3, 1024^3$ and $2048^3$. Top, middle and bottom panels have the values $|f_{R0}| = 10^{-4}, 10^{-5}$ and $10^{-6}$. In any case we use a seven-point gradient operator.}
\label{fig:pk_fr_gravity_emantis}
\end{figure}

In this section, we validate the implementation of the $f(R)$ gravity model from \cite{hu2007models} described in Section~\ref{sec:theory_fr}. To assess this, we run $f(R)$ simulations with varying $f_{R0}$ values, along with a reference Newtonian simulation. Furthermore, because $f(R)$ corrections are irrelevant at very high redshifts, we use the same Newtonian initial conditions in both cases. In Fig.~\ref{fig:pk_fr_gravity_emantis}, we compare the power spectrum boost from \pysco with the \emantis emulator \citep{saez-casares2024emantis}, which is based on the \ecosmog code \citep{li2012ecosmog}, itself a modified version of \ramses.
The results show excellent agreement between \pysco and \emantis up to $k\sim 1~h\mathrm{Mpc}^{-1}$. The agreement improves further for higher $n_{\rm cells}$, except in the case of $|f_{R0}| = 10^{-6}$, where the curves for \pysco overlap when using FFT. This indicates that \pysco converges well towards the \emantis predictions. Notably, the best agreement is found when using the multigrid solver for $|f_{R0}| = 10^{-4}$ and $10^{-5}$, while FFT performs better for $|f_{R0}| = 10^{-6}$. This last behaviour could be explained by the multigrid PM solver struggling to accurately compute small-scale features of the scalaron field, as lower values of $|f_{R0}|$ result in sharper transitions between Newtonian and modified gravity regimes.
No noticeable impact from the gradient order is observed in any of the cases.

In summary, \pysco demonstrates excellent agreement with \emantis, aligning with prior validation efforts against other codes for similar tests, such as those conducted by \cite{adamek2024KPJC6P1}. For this setup, using the multigrid solver for the Newtonian part seems advantageous for consistency, given that its non-linear version is already employed for the additional $f(R)$ field without approximations.

\subsection{Parametrised gravity}

\begin{figure}
\centering
\includegraphics[width=1.0\hsize]{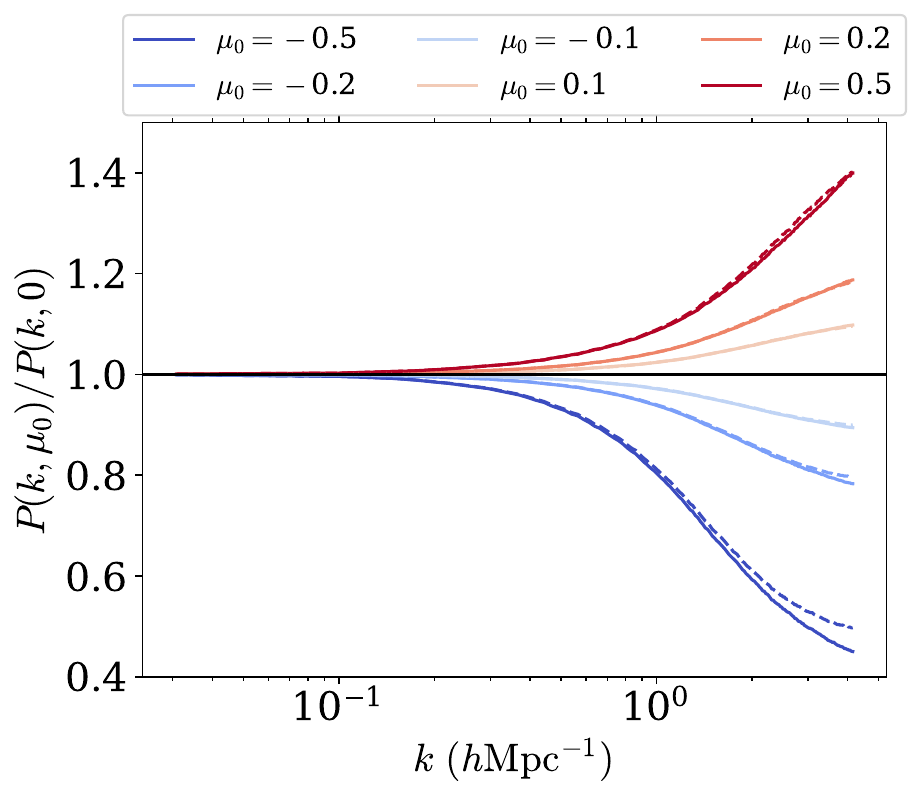}
\caption{Power spectrum boost from parametrised gravity w.r.t to the Newtonian case $(\mu_0 = 0)$. Coloured lines refer to different values of $\mu_0$, while solid and dashed lines indicates the use of FFT and multigrid solvers respectively.}
\label{fig:pk_parametrized_gravity}
\end{figure}

The focus here shifts to simulations with parametrised gravity, where deviations from Newtonian gravity are governed by a single parameter, $\mu_0$, representing the gravitational coupling today (as discussed in Section~\ref{sec:theory_parametrized_gravity}). In Fig.~\ref{fig:pk_parametrized_gravity}, the power spectrum boost is shown for various values of $\mu_0$ compared to a Newtonian simulation.
On large scales, the power spectrum ratio approaches unity, which aligns with expectations since the power spectrum is rescaled at the initial redshift according to $\mu_0$ (detailed in Appendix~\ref{appendix:growth_factors}). On smaller scales, the behaviour changes: negative values of $\mu_0$ result in a suppression of power, while positive values lead to an excess of power. The magnitude of these deviations increases with larger $|\mu_0|$, and the asymmetry between positive and negative values becomes evident. For instance, the departure from Newtonian behaviour is around 60\% for $\mu_0 = -0.5$ and around 40\% for $\mu_0 = 0.5$.

There is a slight discrepancy between the results of the FFT and multigrid solvers for larger values of $|\mu_0|$, although no significant impact from the gradient stencil order or the number of cells is observed at the same scales.

\subsection{MOND}

\begin{figure}
\centering
\includegraphics[width=1.0\hsize]{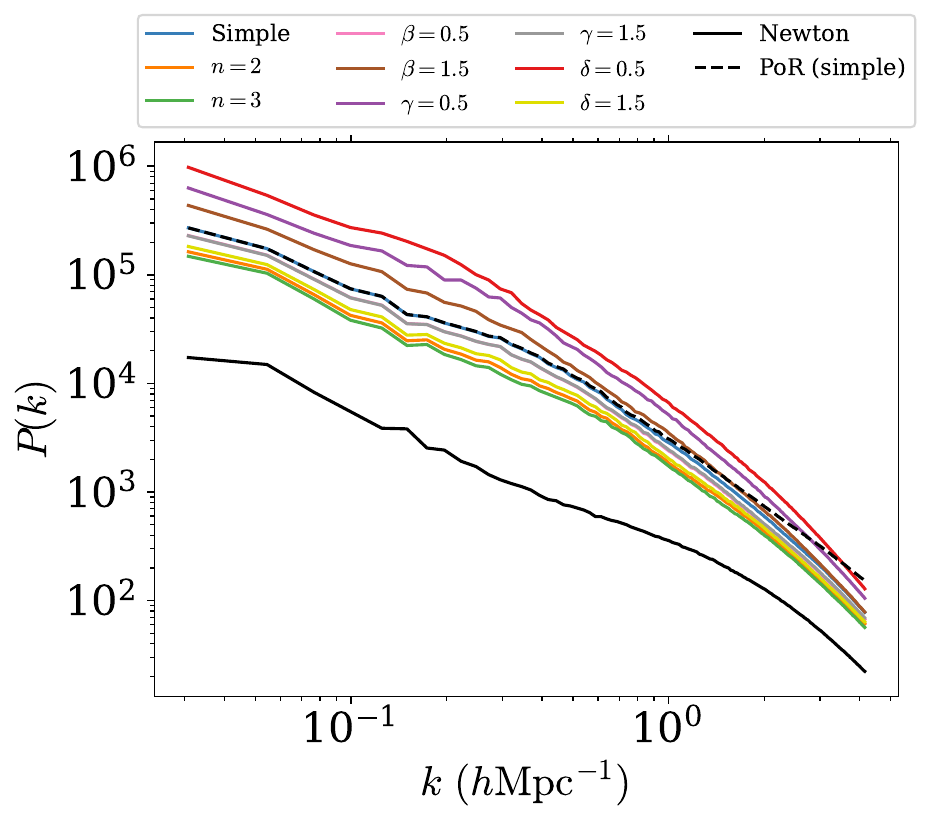}
\caption{Power spectra for several MOND interpolating functions and parameters (in coloured lines). In back solid line we show the Newtonian counterpart, while the black dashed line refers to a MOND run with the simple parametrisation using the code \texttt{Phantom of RAMSES}. For MOND simulations we use $g_0 = 10^{-10}~$m.s$^{-2}$ and $\mathcal{N} = 0$. In \pysco we use the multigrid solver in any case.}
\label{fig:pk_mond}
\end{figure}

This section discusses the testing of the MOND implementation within \pysco. MOND was originally proposed as an alternative explanation for dark matter, modifying Newton's gravitational law in low-acceleration regimes. However, for validation purposes, the same cosmological parameters and initial conditions from Section~\ref{sec:comparison_to_ramses} are used, rather than the typical MOND universe with $\Omega_m = \Omega_b$ as the goal is to test the MOND gravity solver. The results are illustrated in Fig.~\ref{fig:pk_mond}.
The MOND power spectra are noticeably higher than the Newtonian reference. This result aligns with known characteristics of MOND, which accelerates structure formation \citep{sanders2001formation, nusser2002mond}, explaining why MOND simulations are usually initialised with lower values of $A_s$ (or $\sigma_8$) \citep{knebe2004galactic, llinares2008cosmological}. To validate the MOND implementation, it is compared to the \texttt{PoR} (Phantom of \ramses) code \citep{lughausen2015por}, a MOND patch for \ramses that uses the simple interpolating function from Equation~\eqref{eq:mond_simple}. The agreement between \texttt{PoR} and \pysco is excellent for scales $k\lesssim 1~h/$Mpc.

The discrepancies observed at small scales stem from differences between the PM and PM-AMR solvers, a pattern also seen in Fig.~\ref{fig:pk_solvers_ramses_pm}. Furthermore, the impact of the interpolating function on the power spectrum follows the same trend as observed in Fig.~\ref{fig:mond_interpolating_functions}. For consistency, we also verified that MOND power spectra converge towards the Newtonian case when $g_0 \ll 10^{-10}~$m.s$^{-2}$.

\subsection{Performances}
\label{sec:results_performances}

Finally, we present performance metrics for \pysco on the Adastra supercomputer at CINES, using AMD Genoa EPYC 9654 processors. For FFT-based methods, the \texttt{PyFFTW} package\footnote{\faicon{github}~\href{https://github.com/pyFFTW/pyFFTW}{https://github.com/pyFFTW/pyFFTW}.}, a Python wrapper for the FFTW library \citep{frigo2005fftw}, was used. All performance tests were run five times, and the median timings were taken to avoid outliers from potential node-related issues.

\begin{figure}
\centering
\includegraphics[width=1.0\hsize]{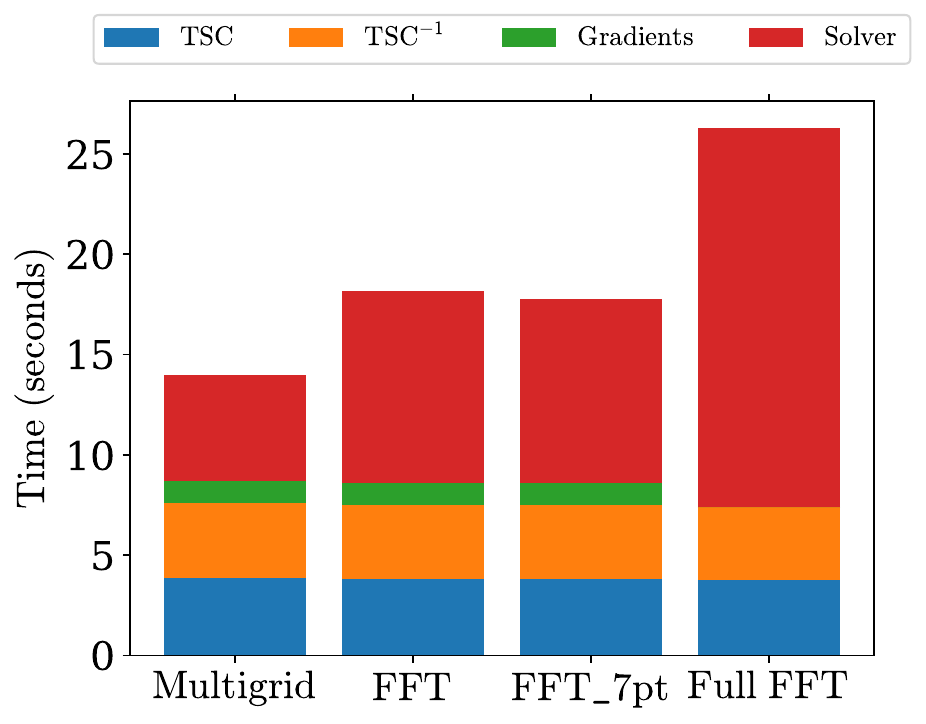}
\caption{Time to compute one time step of a Newtonian simulation for different solvers. In blue, orange, green and red we highlight the time needed to compute a density grid using TSC, an inverse TSC to interpolate the force from grid to particles, a seven-point gradient and the solver respectively. The FULL\_FFT method directly computes the force with the kernel described in Eq.~\eqref{eq:exact_fullfft_kernel}, hereby removing the need for a finite-derivative gradient part. All timings were done using a single CPU.}
\label{fig:scaling_solvers}
\end{figure}

The first benchmark focuses on the time required to compute a single time step, as shown in Fig.~\ref{fig:scaling_solvers}.
In a simulation with $512^3$ particles and cells, the computation takes between 15 to 25 seconds on a single CPU for the various solvers, with the multigrid solver being the fastest and the FULL\_FFT solver being the slowest. The force computation using a seven-point gradient from the gravitational potential grid contributes minimally to the overall runtime. Since the multigrid solver outperforms FFT\_7pt with similar accuracy, the latter will not be used in the remainder of the paper.

\begin{figure}
\centering
\includegraphics[width=1.0\hsize]{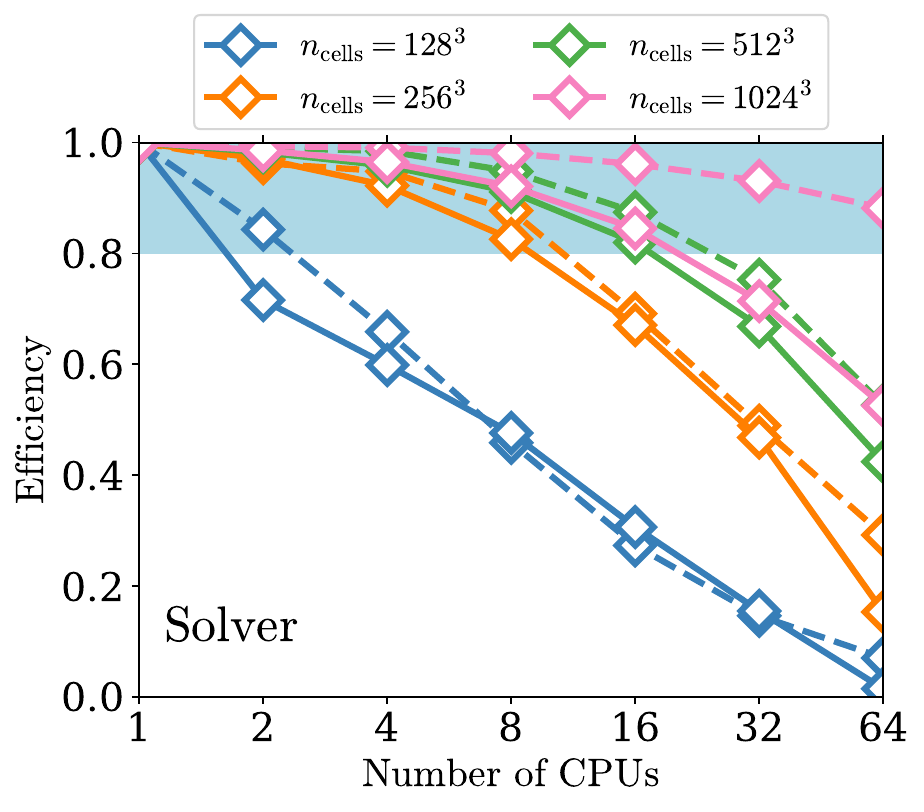}
\includegraphics[width=1.0\hsize]{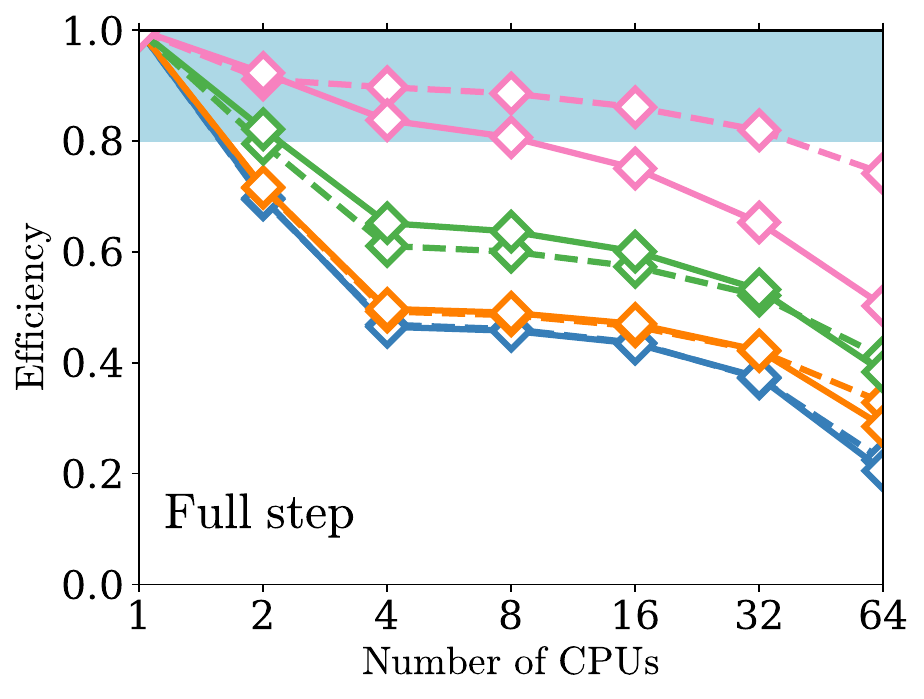}
\caption{Strong scaling efficiency for the solver only (top panel) and full time step (bottom panel). The solid and dashed lines refer to FFT and multigrid solvers respectively. The efficiency is computed as the time ratio for one CPUs w.r.t $N$ CPUs, divided by $N$. An efficiency equal to unity denotes perfect parallel scaling.}
\label{fig:scaling_strong}
\end{figure}

Fig.~\ref{fig:scaling_strong} illustrates the strong scaling efficiency of \pysco's FFT and multigrid solvers.
The scaling improves as the number of cells increases, suggesting that the workload per CPU becomes more efficient with larger grids. For smaller grids ($n_{\rm cells} = 128^3$), the multithreading is less effective, whereas for larger grids ($n_{\rm cells} = 1024^3$), multigrid reaches an efficiency of roughly 90\% with 64 CPUs. Overall, multigrid consistently exhibits better efficiency than FFTW

When analysing the total time per time step, a slightly different picture emerges. Efficiency still improves with larger grids but is generally lower than solver-only performance. For smaller grids ($n_{\rm cells} = 128^3$ and $256^3$), there is little difference between multigrid and FFT, indicating that particle-grid interactions are the primary factor influencing performance, as $512^3$ particles were used in all cases. A significant drop in efficiency occurs when using fewer than four CPUs, which is followed by a more gradual decline. This is caused by race conditions in the TSC algorithm, where multiple threads attempt to write to the same element in the density grid. To address this, atomic operations were implemented to ensure thread-safe modifications, but these operations slow the mass-assignment process by a factor of four\footnote{Unfortunately, \numba does not currently contain a way to deal with race conditions on CPUs natively. We thus had to rely on a modified version of atomic operations developed in \href{https://github.com/KatanaGraph/katana}{https://github.com/KatanaGraph/katana}. We will move on the official implementation of \numba atomics once they are developed.}. Hence, when $N_{\rm CPU} < 4$ we use the sequential TSC version (which thus does not scale at all by definition), and the parallel-safe version for $N_{\rm CPU} \geq 4$, thus giving a better scaling afterwards.

For grids with $n_{\rm cells} = 512^3$, FFT achieves better efficiency than multigrid because, while multigrid is more efficient as a solver, it constitutes a smaller portion of the overall runtime. However, with larger grids $n_{\rm cells} = 1024^3$, the efficiency aligns more closely with the solver-only case, as the solver dominates the runtime. Efficiency reaches approximately 50\% for FFT and 75\% for multigrid. To ensure optimal efficiency, the number of grid cells should thus be at least eight times the number of particles  $n_{\rm cells} \geq 8 n_{\rm part}$.

For comparison, a full simulation for \pysco~with 512$^3$ particles and as many cells takes roughly 0.8 CPU hours for $\sim$200 time steps, while a \ramses~run with the same setup and AMR enabled takes roughly 3000 CPU hours for $\sim$1000 time steps.

\section{Conclusions}
\label{sec:conclusions}
In this paper, we presented \pysco, a fast particle-mesh $N$-body code designed for Newtonian and modified-gravity cosmological simulations. \pysco currently supports the $f(R)$ gravity model \citep{hu2007models}, the quasi-linear formulation of MOND \citep{milgrom2010quasilinear}, and a time-dependent modification of the gravitational constant via a parameterised model. The code includes multiple solvers, such as multigrid and several FFT methods, each with exact or approximated kernels. We validated \pysco against \ramses (using multigrid and no AMR), the \emantis emulator for $f(R)$ gravity and \texttt{Phantom of RAMSES} for MOND, and found overall good agreement.
The main conclusions of our study are the following:
\begin{itemize}
    \item The TSC mass assignment should be preferred over CIC. CIC leads to larger scatters in the power spectrum and inaccuracies, particularly with FFT solvers when the number of cells exceeds the number of particles.
    \item Three-point gradient operators result in a loss of power at small scales. Using at least a five-point gradient is recommended for better accuracy.
    \item Seven-point Laplacian operators, whether used in multigrid (configuration space) or FFT\_7pt (Fourier space), cause power suppression at small scales. Multigrid is favoured because it provides undistinguishable results from FFT\_7pt, while being faster and more efficient in parallel computing. 
    \item The FFT solver (using the exact Laplacian kernel) is a solid choice overall, especially with TSC instead of CIC when the number of cells exceeds the number of particles.
    \item The FULL\_FFT solver is too simplistic to be trusted, except in configurations where the number of cells matches the number of particles ($n_{\rm cells} = n_{\rm part}$) and TSC is used.
    \item Resolving small scales sufficiently is critical for obtaining a power spectrum at $z = 0$ that is relatively insensitive to the initial redshift. Otherwise, there will be bias at small scales. Solutions include increasing the resolution (for example, reducing box size), using a higher-order gradient operator, or applying an exact Laplacian operator.
    \item When small scales are well-resolved, the power spectrum at $z = 0$ becomes insensitive to the starting redshift at around 0.1\% for initial redshifts greater than 125, 30, and 10 for 1LPT, 2LPT, and 3LPT, respectively. 
    \item We recommend running \pysco with $n_{\rm cells} = 8 n_{\rm part}$ for optimal efficiency. Increasing to $n_{\rm cells} = 64 n_{\rm part}$ would significantly raise the runtime, contradicting the main advantage of PM-only simulations, which is precisely to be faster than AMR or tree codes.
    \item For standalone simulations, the FFT solver is ideal, as it aligns best with PM-AMR results. For calculating boost factors, such as $f(R)$ simulations against Newtonian ones, multigrid is a faster option, with power losses at small scales cancelling out.
\end{itemize}
In this paper we focused on power spectrum, but it would be also possible to compute accurate boost factors for halo mass functions (HMF), as in \cite{adamek2024KPJC6P1}, although the HMF itself (not the boost) is very sensitive to small scales.

\pysco is usable both through command line and as a Python package. It also includes a wide range of built-in utilities, such as initial condition generators and power spectrum estimators. To minimise third-party dependencies, PySCo integrates several in-house modules for particle and mesh algorithms.


In the future, we will port \pysco to GPUs using \texttt{Numba} for CUDA kernels. However, there is currently no plan to implement MPI parallelisation, as the typical PM use case involves small, fast simulations on a single node.
\pysco is particularly interesting because it comes with the fast development speed of Python and C/Fortran-like performance.
Compared to PM-AMR codes like \ramses, \pysco offers a dramatic reduction in runtime—potentially simulating in roughly 1/1000th of the time, albeit with less accuracy on small scales. These features make it useful for producing emulators, generating covariance matrices, and training neural networks for upcoming cosmological surveys without requiring massive computational resources. Our aim is to expand \pysco to include a wider range of gravity theories and non-standard dark sector components, creating a versatile framework for exploring cosmological phenomena and aiding in the development of observational constraints.


\begin{acknowledgements}
I would like to thank Yann Rasera,  I\~nigo S\'aez-Casares and Himanish Ganjoo for comments on the draft and testing a preliminary version of \pysco. I am also grateful to Stéphane Colombi and Oliver Hahn for insightful discussions regarding initial conditions, as well as Hans Winther on modified gravity simulations and Claudio Llinares for interesting prospects with MOND. I thank Jorge Carretero and Pau Tallada, whose work on the \texttt{SciPic} package convinced me of the viability of Python for performant scientific computing.

This project was provided with
computer and storage resources by GENCI at CINES thanks to the grant 2023-A0150402287 on the supercomputer Adastra’s GENOA partition.

\end{acknowledgements}

\bibliographystyle{aa}
\bibliography{biblio} 

\appendix

\section{Growth factors}
\label{appendix:growth_factors}

Under the Lagrangian formulation of perturbation theory, the position of tracers at some given time and position is given by 
\begin{equation}
    \bm{x}(\eta) = \bm{q} + \Psi(\bm{q}, \eta),
\end{equation}
with $\bm{q}$ an initial position and $\Psi(\bm{q}, \eta)$ a displacement field. The Euler equation is then
\begin{equation}
    \frac{\dd^2\bm{x}}{\dd\eta^2} + \mathcal{H}\frac{\dd\bm{x}}{\dd\eta} = -\nabla\phi,
\end{equation}
which, when taking the gradient, yields
\begin{equation}
   \nabla_{\bm{x}}\left[\frac{\dd^2\bm{x}}{\dd\eta^2} + \mathcal{H}\frac{\dd\bm{x}}{\dd\eta}\right] = -\frac{3}{2}\mathcal{H}^2\Omega_m(\eta)\delta(\bm{x}, \eta).
\end{equation}
Mass conservation imposes $\bar{\rho}(\eta)\dd^3\bm{q} = \bar{\rho}(\eta)\left[1+\delta(\bm{x},\eta)\right]\dd^3\bm{x}$, meaning that $\nabla_{\bm{x}} = J^{-1}\nabla_{\bm{q}}$, with $J(\bm{q},\eta) = \mathrm{det}\left[\delta_{ij}+\Psi_{ij}(\bm{q}, \eta)\right]$ where $\delta_{ij}$ is a Kronecker Delta. We compute the time evolution of growth factors through the general equation \citep{jeong2010cosmology}
\begin{equation}
\label{eq:general_equation_growth}
    J\nabla_{\bm{x}} \left[\frac{\dd^2\bm{x}}{\dd\eta^2} + \mathcal{H}\frac{\dd\bm{x}}{\dd\eta}\right] = \frac{3}{2}\mathcal{H}^2\Omega_m(\eta)\left[J - 1\right].
\end{equation}
A perturbative expansion at third order gives \citep{rampf2012lagrangian}
\begin{align}
    \Psi(\bm{q},\eta) &= \varepsilon D(\eta)\Psi^{(1)}(\bm{q}) + \varepsilon^2 E(\eta)\Psi^{(2)}(\bm{q}) + \varepsilon^3 F(\eta)\Psi^{(3)}(\bm{q}) + \mathcal{O}(\varepsilon^4), \\
    J(\bm{q}, \eta) &= 1 + \varepsilon D(\eta)\mu_1^{(1)}(\bm{q}) + \varepsilon^2\left[E(\eta)\mu_1^{(2)}(\bm{q}) + D^2(\eta)\mu_2^{(1)}(\bm{q})\right] \nonumber\\
    & + \varepsilon^3 \left[F(\eta)\mu_1^{(3)}(\bm{q}) + 2 D(\eta)E(\eta)\mu_2^{(1,2)}(\bm{q}) + D^3(\eta)\mu_3^{(1)}(\bm{q})\right] \nonumber \\
    &+ \mathcal{O}(\varepsilon^4),
\end{align}
with $D \equiv D^{(1)}$, $E \equiv D^{(2)}$ and $F \equiv D^{(3)}$ are the first, second and third-order growth factors respectively, $\varepsilon \ll 1$ and
\begin{align}
    &\mu_1^{(n)}(\bm{q}) = \Psi_{i,i}^{(n)}(\bm{q}), \\
    &\mu_2^{(n,m)}(\bm{q}) = \frac{1}{2}\left[\Psi_{i,i}^{(n)}(\bm{q})\Psi_{j,j}^{(m)}(\bm{q}) - \Psi_{i,j}^{(n)}(\bm{q})\Psi_{j,i}^{(m)}(\bm{q})\right], \\
    &\mu_3^{(n)}(\bm{q}) = \mathrm{det}\left[\Psi_{i,j}^{(n)}(\bm{q})\right].
\end{align}
Eq.~\eqref{eq:general_equation_growth} then becomes
\begin{equation}
    J\left[\delta_{ij} - \Psi_{i,j}\right]\left[\frac{\dd^2\Psi_{i,j}}{\dd\eta^2} + \mathcal{H}\frac{\dd\Psi_{i,j}}{\dd\eta}\right] = \frac{3}{2}\mathcal{H}^2\Omega_m(\eta)\left[J - 1\right],
\end{equation}
with $\partial/\partial x_i = \left[\delta_{ij} + \Psi_{i,j}\right]^{-1}\partial/\partial q_j$ and $\left[\delta_{ij} + \Psi_{i,j}\right]^{-1} \approx \left[\delta_{ij} - \Psi_{i,j}\right]$.

- The first-order solution \citep{zeldovich1970gravitational} is
\begin{equation}
    \ddot{D} + \mathcal{H}\dot{D} -\beta D = 0,
\end{equation}
with $\beta = \frac{3}{2}\mathcal{H}^2\Omega_m(\eta)$.
where $D \equiv D_+$ the first-order growth factor, and $\dot{D} = \dd D/\dd\eta$.

- The second-order solution is then
\begin{equation}
    \ddot{E} + \mathcal{H}\dot{E} -\beta\left[E - D^2\right] = 0.
\end{equation}

- The third-order solutions are given by
\begin{align}
    &\left[\ddot{F} + \mathcal{H}\dot{F} - \beta F\right]\mu_1^{(3)} \nonumber\\
    & + 2\left[D \ddot{E} + \mathcal{H}D\dot{E} + \ddot{D}E\ + \mathcal{H}E\dot{E} - \beta DE\right]\mu_2^{(1,2)}  \nonumber\\
    & + \left[3\ddot{D}D^2 + 3\mathcal{H}\dot{D}D^2-\beta D^3\right]\mu_3^{(1)} = 0,
\end{align}
which give rise to two tangential and one transversal modes. Their ordinary differential equations (ODEs) are given by
\begin{align}
&\ddot{F}_a + \mathcal{H}\dot{F}_a  - \beta\left[F_a - 2D^3\right] = 0, \\
&\ddot{F}_b + \mathcal{H}\dot{F}_b  - \beta\left[F_b - D\left(E - D^2\right)\right] = 0, \\
&\ddot{F}_c + \mathcal{H}\left[\dot{E}D - E\dot{D}\right]  + \beta D^3 = 0,
\end{align}
where the last equation can be derived using the equation for irrotational fluids \citep{catelan1995lagrangian}.
In practice, we implement these ODEs as functions of $\ln(a)$, leading to the following set of equations
\begin{align}
D'' &= -\gamma D' + \beta D, \\ 
E'' &= -\gamma E' + \beta \left[E - D^2\right], \\
F_a'' &= -\gamma F_a' + \beta \left[F_a - 2D^3\right], \\
F_b'' &= -\gamma F_b' + \beta \left[F_b - 2D\left(E - D^2\right)\right], \\
F_c'' &= (1-\gamma) F_c' + E D' - D E' - \beta D^3, 
\end{align}
with $D' = \dd D/\dd\ln{a}$ and
\begin{align}
\beta &= \frac{3}{2}\Omega_m(a), \\
\gamma &= \frac{1}{2}\left\{1 - 3\Omega_\Lambda(a)\left[w_0 + w_a(1-a)\right] - \Omega_r(a)\right\}. 
\end{align}
The $n$th-order growth factors can be estimated through $f^{(n)} = D^{(n)}{}'/D^{(n)}$.
It is straightforward to implement a scale-independent parametrised form of modified gravity (as in Section~\ref{sec:theory_parametrized_gravity}) by making the small change $\beta = 3\mu(a)\Omega_m(a)/2$.

Approximated analytical solutions for the growing mode in a matter-dominated era (and neglecting radiation) are \citep{catelan1995lagrangian, rampf2012lagrangian}
\begin{align}
D_+^{(2)} &=  -\frac{3}{7}D_+^2, \\
D_+^{(3a)} &= -\frac{1}{3} D_+^3, \\
D_+^{(3b)} &= \frac{10}{21}D_+^3, \\
D_+^{(3c)} &= -\frac{1}{7}D_+^3.
\end{align}
Approximated fits with dependence on $\Omega_m$ for the growth factors and growth rates are also available in \cite{bouchet1995perturbative}.

\section{Initial conditions}
\label{appendix:initial_conditions}

\subsection{Time stepping}
\label{appendix:time_stepping}

As described in Section~\ref{sec:integrator}, we use two kinds of time stepping criteria, 
one based on cosmological time (in \ramses, the scale factor cannot change by more than 10\% in a single time step), and another based on the free-fall time (acceleration) or maximum velocity compared to the mesh size. For the latter two we also use a Courant-like factor, which multiplies the final time step and is equal to 0.8 by default.

\begin{figure}
\centering
\includegraphics[width=1.0\hsize]{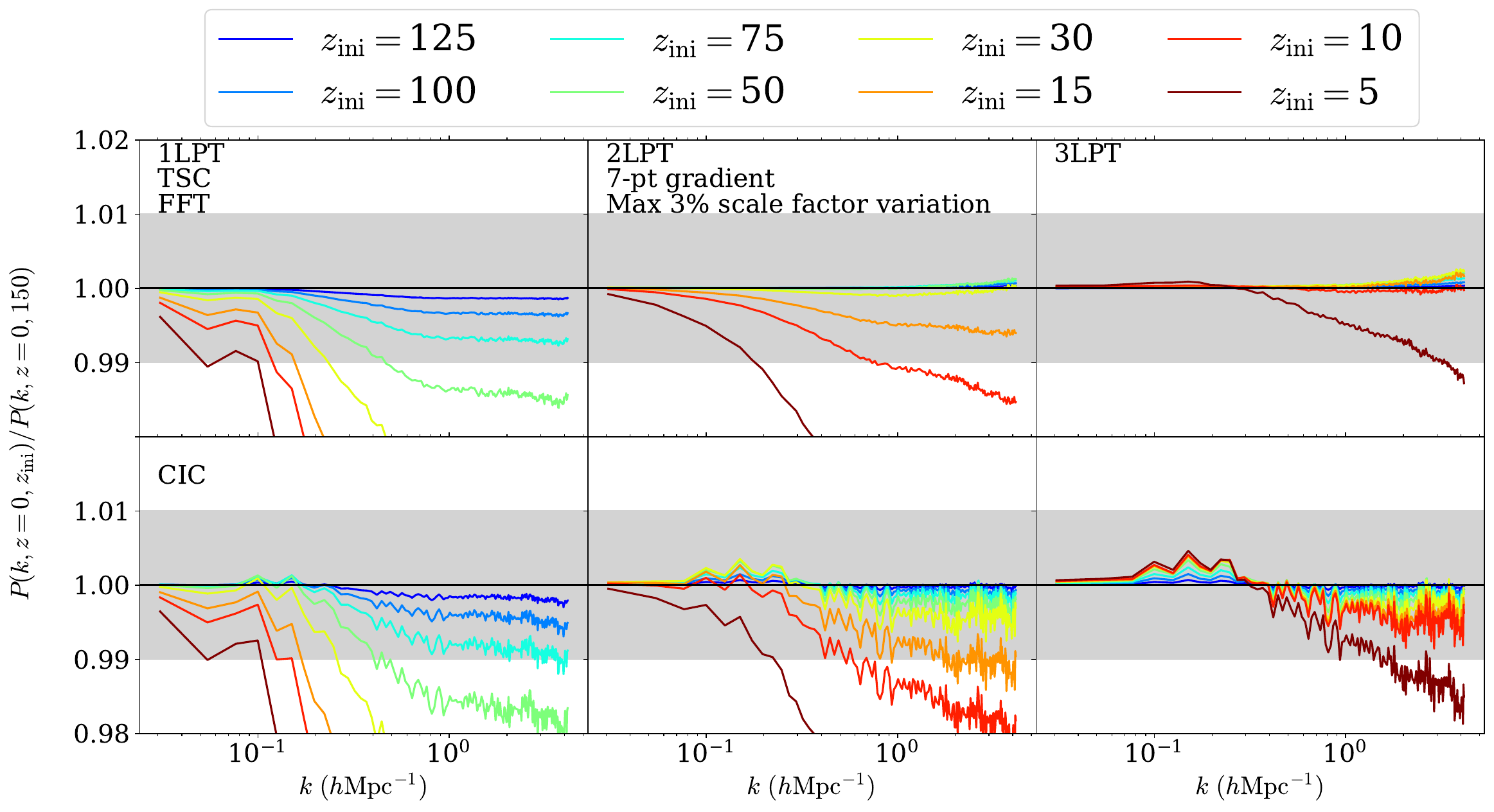}
\includegraphics[width=1.0\hsize]{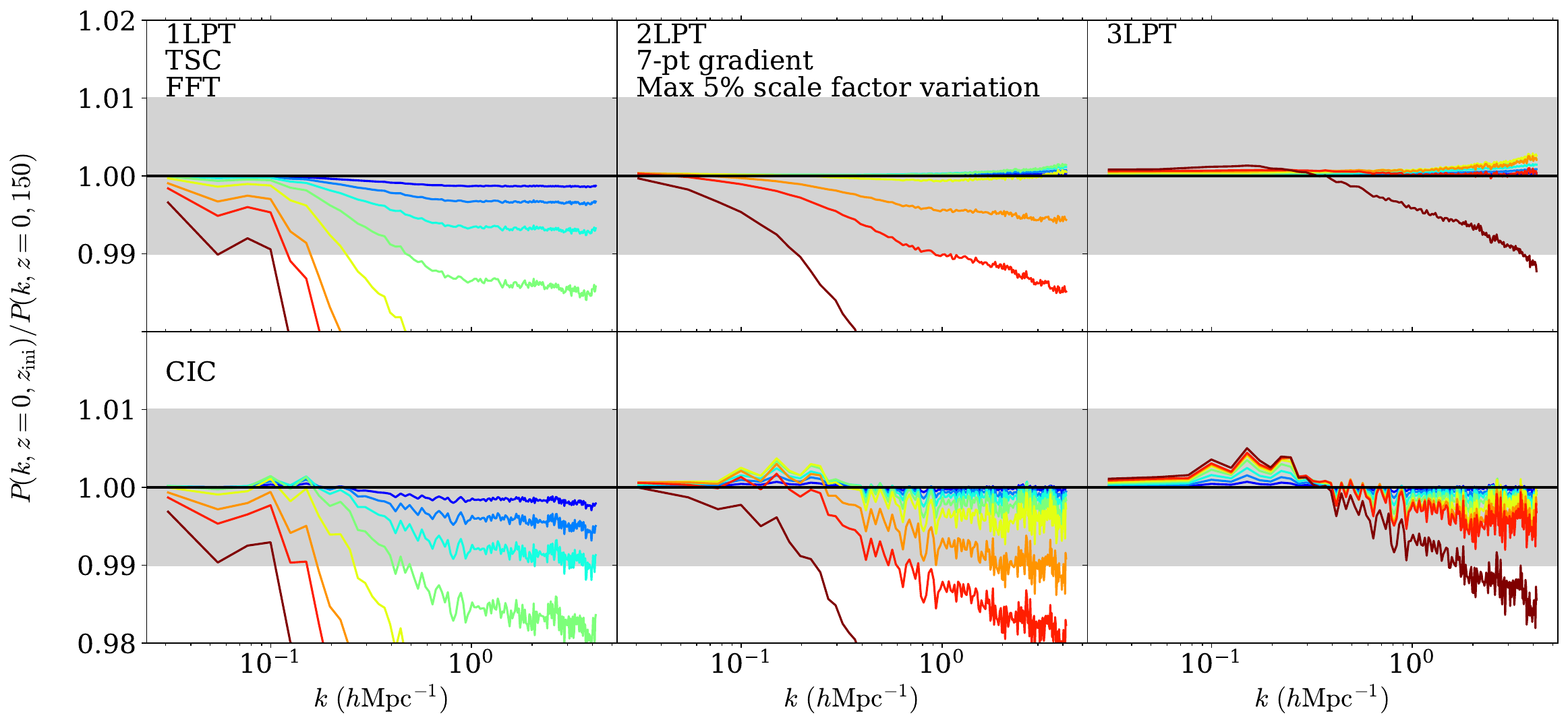}
\includegraphics[width=1.0\hsize]{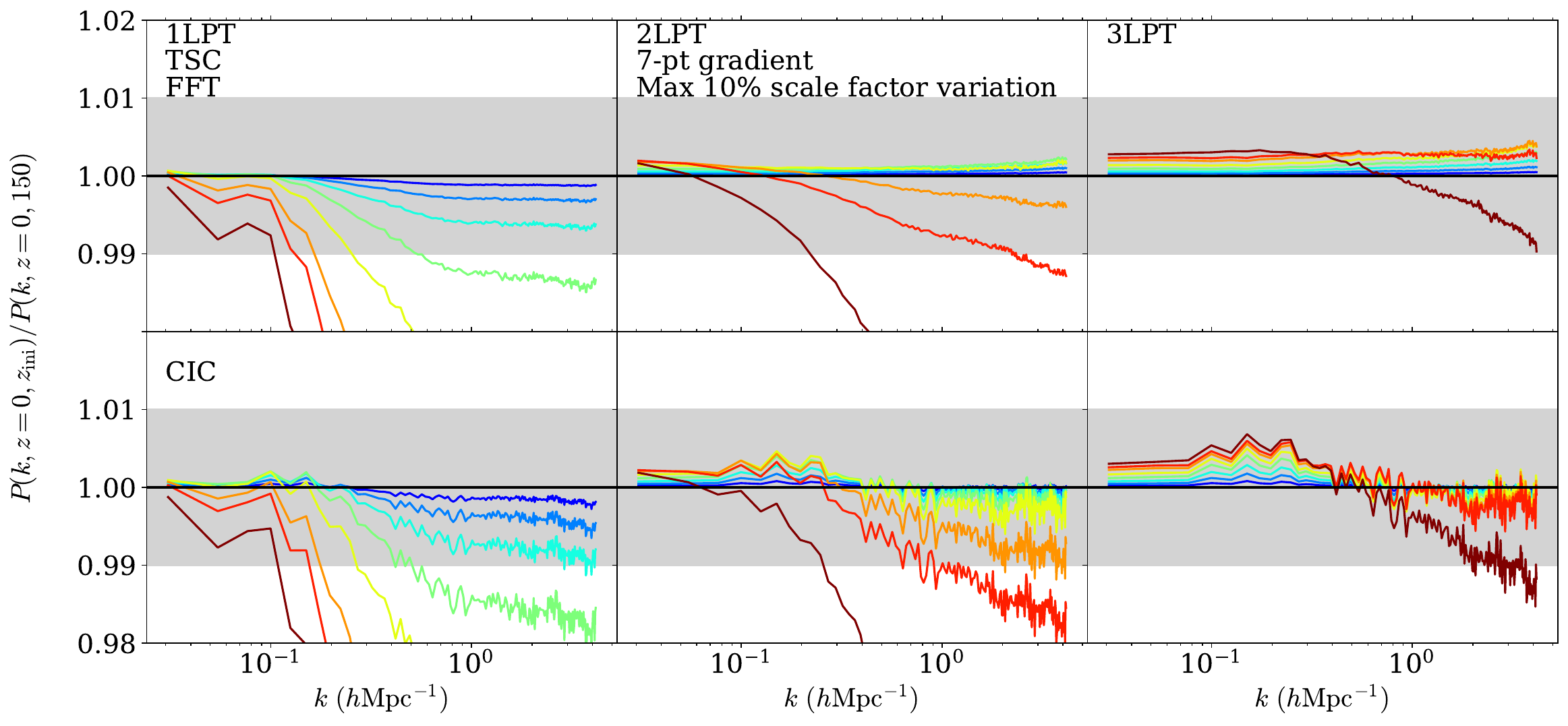}
\caption{Same as Fig.~\ref{fig:initial_conditions_gradients}, but using a seven-point gradient and varying the maximal scale factor variation in the cosmological time criterion, from top to bottom.}
\label{fig:initial_conditions_cosmo_time}
\end{figure}

While small changes in the Courant factor do not significantly affect the results, using a time step that is too large can introduce a scale-independent bias in the power spectrum at redshift $z=0$, as shown in Fig.~\ref{fig:initial_conditions_cosmo_time}. 
We observe a bias of approximately 0.3\% in the power spectrum at $k \sim 0.1~h/$Mpc when the time step allows for a 10\% maximum variation in the scale factor. This bias decreases to about 0.1\% with a 5\% variation and seems negligible when using variations below 3\%.
To ensure accuracy and avoid such biases, we used a more conservative 2\% variation in scale factor by default. Additionally, we note that while 1LPT appears to work well on large scales, this is primarily due to the limited box size of the simulation, as we directly see the moment when the power spectrum is damped compared to the reference at initial redshift $z = 150$.

\subsection{Initial positions}
\label{appendix:initial_positions}

\begin{figure}
\centering
\includegraphics[width=1.0\hsize]{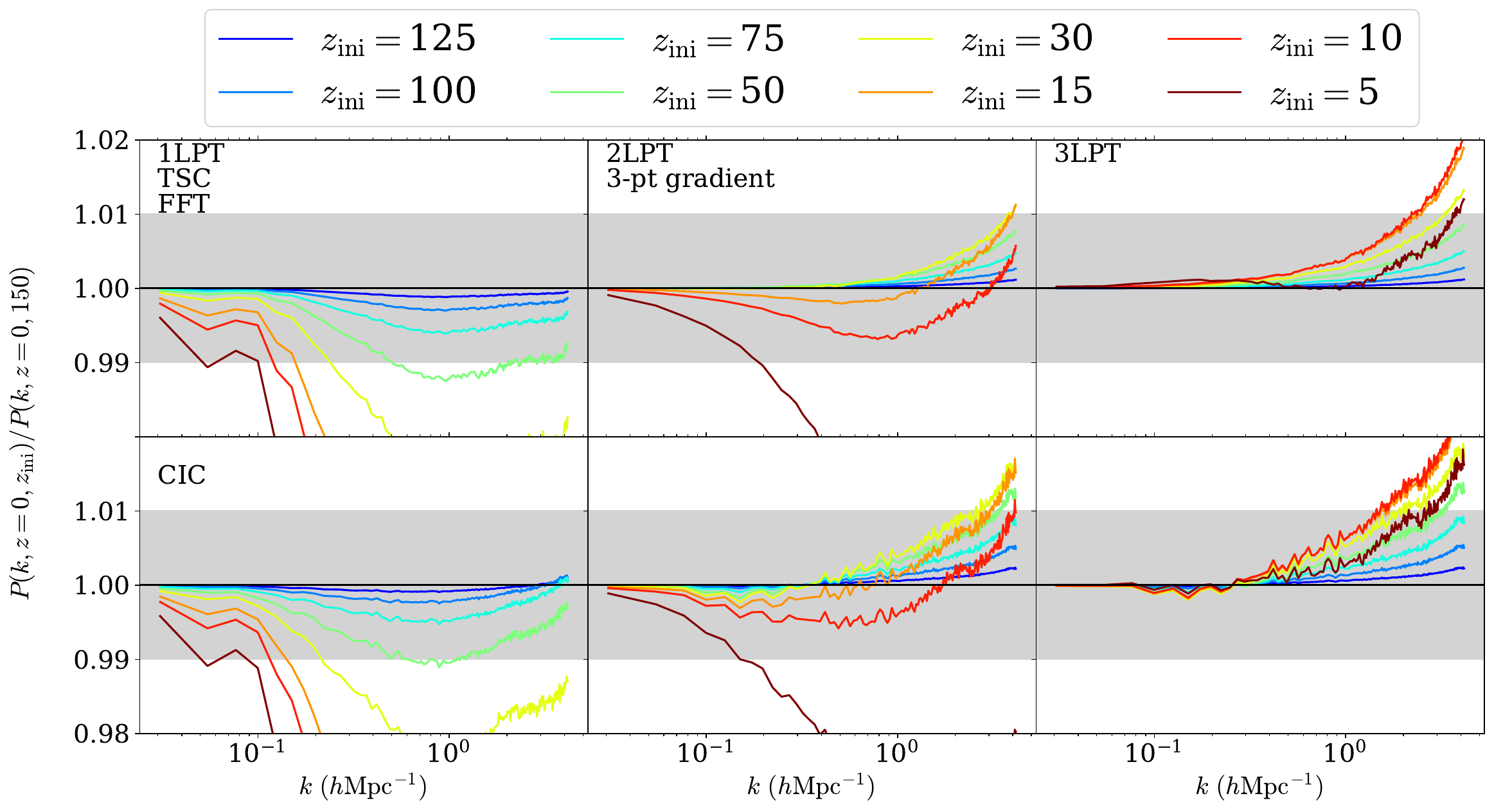}
\includegraphics[width=1.0\hsize]{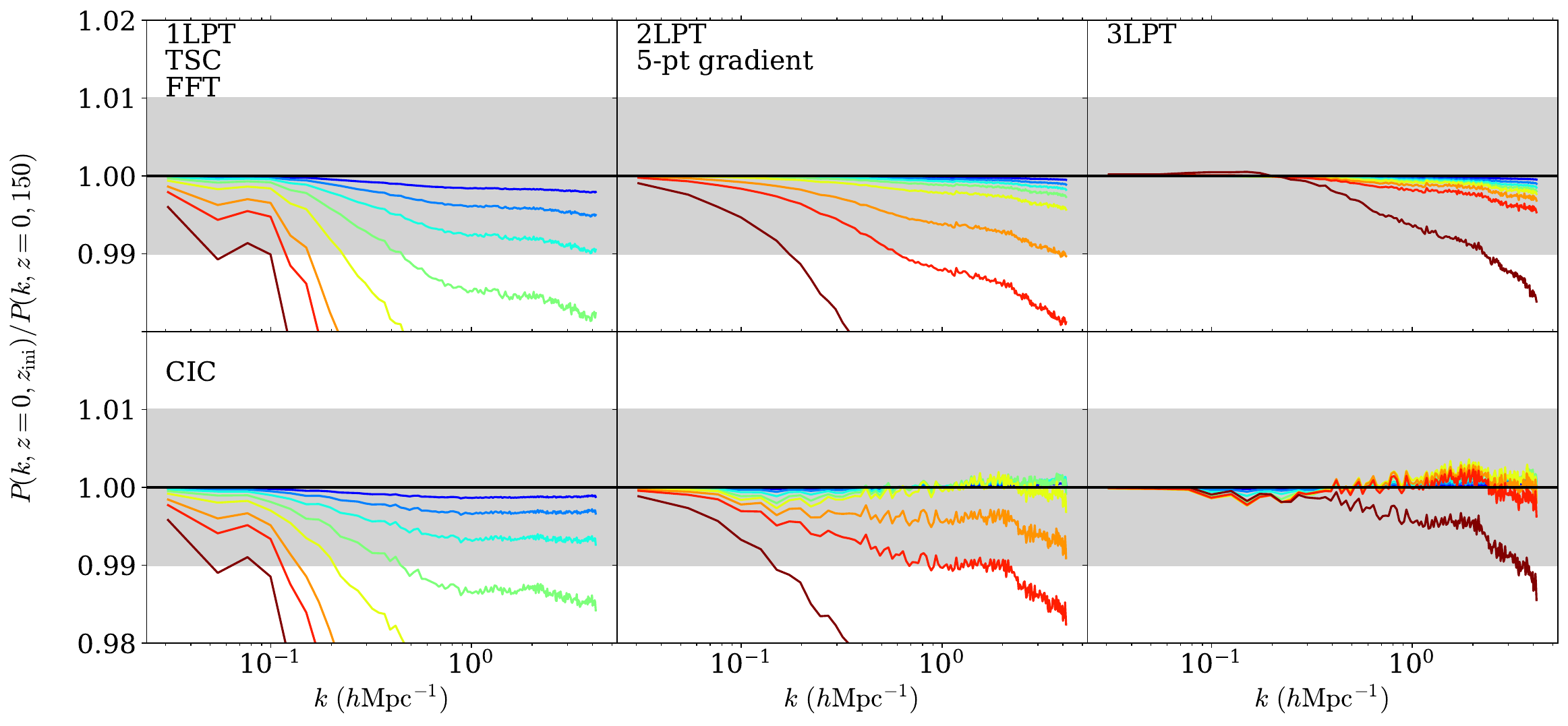}
\includegraphics[width=1.0\hsize]{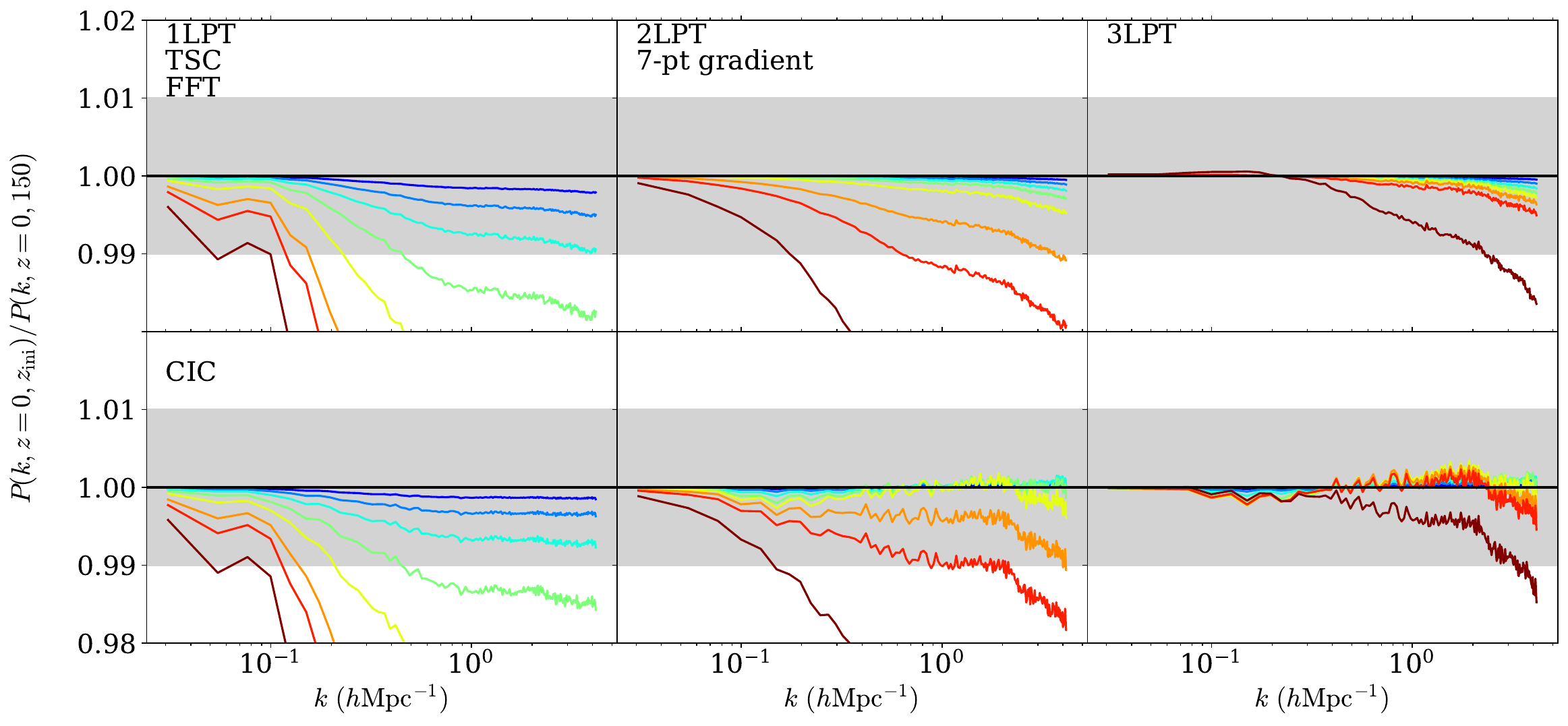}
\caption{Same as Fig.~\ref{fig:initial_conditions_gradients}, but initialising the particles at cell edges.}
\label{fig:initial_conditions_gradients_edges}
\end{figure}

In \pysco, two methods for initialising particle positions were implemented: at cell centres or at cell edges. Given the periodic boundary conditions, these methods should ideally produce no differences due to translational symmetry. However, discrepancies can arise depending on the mass-assignment scheme used. For example, using the nearest grid point (NGP) method, where the density is calculated directly from the number of particles within each grid cell, will produce a uniform density grid if the particles are positioned at cell centres. This is because, at small displacements, each cell would contain exactly one particle.
In contrast, if the particles are initialised at the cell edges, the density field can become inhomogeneous, as it depends more strongly on the displacement field. This difference is demonstrated in Fig.~\ref{fig:initial_conditions_gradients_edges}, which shows the ratio of the power spectrum for various initial redshifts compared to a reference case where the initial redshift is $z_{\rm ini}=150$ and particles are initialised at cell edges
The results are quite similar to those from Fig.~\ref{fig:initial_conditions_gradients}, where particles were initialised at cell centres. However, for five- and seven-point gradient operators with 3LPT, a slight bias appears at small scales, with a loss of power even for $z_{\rm ini} >10$ compared to the reference case.
As a result of these findings, we decided to initialise particles at cell centres in the main text.

\section{Multigrid convergence}
\label{appendix:multigrid_convergence}
\begin{figure}
\centering
\includegraphics[width=1.0\hsize]{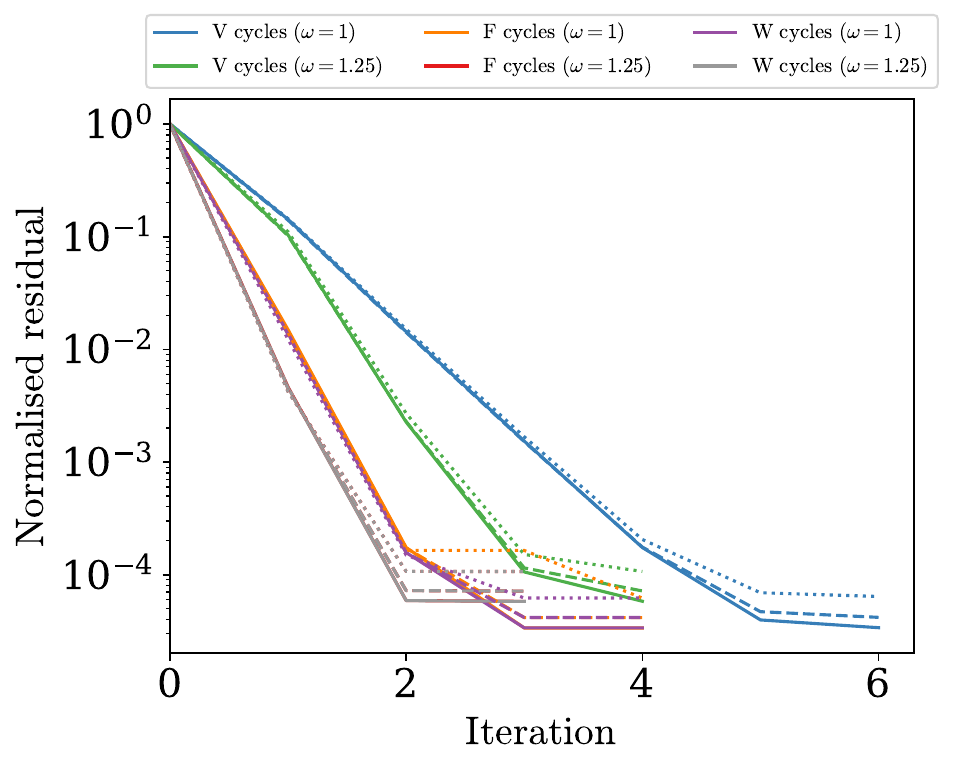}
\caption{Convergence rate of multigrid cycles as function of the number of cycles (or iterations). Blue, orange and purple lines refer to V, F and W cycles respectively, while in green, red and grey we use SOR with relaxation parameter $\omega = 1.25$. In solid, dashed and dotted lines we show results at $z = 0, 1$ and 3 respectively. For the first guess we use one Jacobi step, thereby neglecting the information from previous step.}
\label{fig:multigrid_cycles}
\end{figure}

This section presents convergence tests for the multigrid algorithm (detailed in Section~\ref{sec:multigrid}). Fig.~\ref{fig:multigrid_cycles} shows the convergence rate of different multigrid cycles (V, F, and W).
First, we remark that the convergence rate is nearly independent of redshift. The only redshift-dependent variation is the maximum residual suppression due to numerical roundings, which becomes evident after many iterations. This discrepancy arises from the normalisation by the first-guess residual, which is less accurate at lower redshifts due to the Universe being more structured compared to higher redshifts.
Additionally, F and W cycles show very similar behaviour but take approximately 1.9 and 2.1 times longer than V cycles, respectively. Without overrelaxation ($\omega = 1$), F and W cycles reduce the residual by a factor of 100 per iteration, while V cycles achieve a factor of 10 reduction. Overrelaxation with $\omega = 1.25$ significantly improves the convergence rate, although it may not immediately help in the first V cycle due to the initial guess still being inaccurate. In practice, $\omega = 1.25$ is applied for linear Poisson equations, whereas for non-linear cases, $\omega = 1.0$ is used since SOR proves less effective.
In Fig.~\ref{fig:multigrid_cycles}, one Jacobi step was used as the first guess (see Section~\ref{sec:jacobi_gs_methods}), but Table ~\ref{tab:residuals} suggests that using the potential field from the last step can significantly reduce the initial residuals.  
\begin{table}
  \caption{Ratio of residuals for different first guesses.}
    \smallskip
  \label{tab:residuals}
    \smallskip
  \begin{tabular}{|c|c|c|c|}
    \hline
    \rowcolor{blue!5}
    & & &  \\[-8pt]
    \rowcolor{blue!5}
    Redshift  & No first guess & Last step & Last step + rescale  \\
    \hline
        & & &  \\[-9pt]
    0 & 9.09 & 1.01 & 1  \\
        & & & \\[-9pt]
    \hline
        & & &\\[-9pt]
    1 & 5.99 & 1.02 & 1\\
        & & & \\[-9pt]
    \hline
        & & &\\[-9pt]
    3 & 7.81 & 1.11 & 1 \\
    \hline
  \end{tabular}
  \tablefoot{`No first guess' means that we initialise the potential with one Jacobi step (see Section~\ref{sec:jacobi_gs_methods}), `last step' means that we use the potential field computed at the last step has first guess, and `last step + rescale' has an additional rescaling (as described in Section~\ref{sec:newtonian_simulations}). The latter is used as reference.}
\end{table}
Further improvements through rescaling this last-step solution using the linear growth factor (discussed in Section~\ref{sec:newtonian_simulations}) only offer marginal reductions in the residual. Rescaling becomes more useful at higher redshifts, where time steps are larger. Overall, using the last step as the first guess is a good strategy, sometimes saving one V cycle, though it requires keeping the gravitational potential grid in memory. Rescaling, however, seems unnecessary unless the time steps are particularly large.

\begin{figure}
\centering
\includegraphics[width=1.0\hsize]{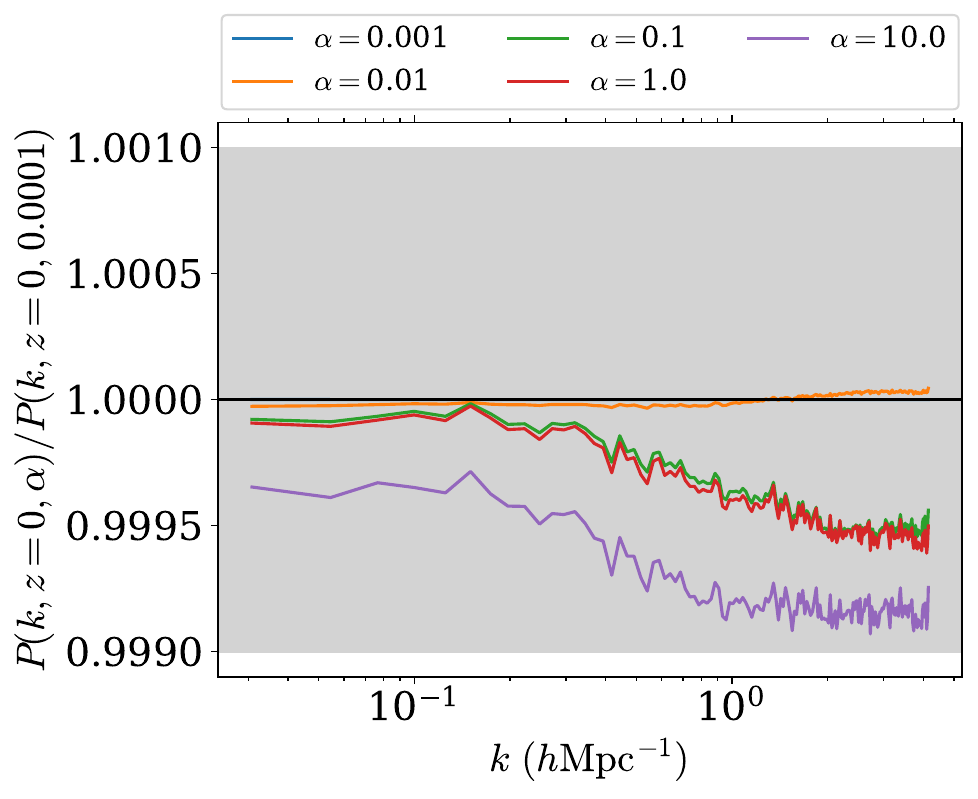}
\caption{Power spectrum ratio with varying multigrid threshold parameter $\alpha$ (see Eq.~\ref{eq:multigrid_threshold}) w.r.t to the reference where $\alpha = 10^{-4}$. In blue, orange, green, red and purple we have $\alpha = 0.001, 0.01, 0.1, 1$ and 10. The grey shaded area indicates the $\pm 0.1\%$ limits. In any case we use V cycles.}
\label{fig:pk_ratio_multigrid_convergence}
\end{figure}

The final test concerns the threshold criterion for convergence of the multigrid algorithm. Fig.~\ref{fig:pk_ratio_multigrid_convergence} shows the power spectrum ratio at $z=0$ for different values of $\alpha$.
Even with a very large threshold value, like $\alpha = 10$, the bias in the power spectrum is only around 0.1\%. The results for $\alpha = 0.1$ and 1 are almost identical, likely because these thresholds are passed within the same V cycle, which typically reduces the residual by a factor of 20–30 per iteration when using SOR. With $\alpha = 0.01$, the bias drops to less than 0.01\%, and convergence is essentially achieved by $\alpha = 0.001$. These results are consistent regardless of starting redshift, LPT order, or mass-assignment scheme.
In practice, an $\alpha$ value of 0.01 is recommended, as it produces highly accurate results without introducing any significant bias, whereas increasing $\alpha$ further results in a small but noticeable decrease in the power spectrum.

\section{Point-mass test}
\label{appendix:point_mass_test}

In this appendix, we evaluate the performance of \pysco~in the simplified scenario of a single point mass while maintaining periodic boundary conditions. The normalised gravitational field for a homogeneous and isotropic sphere of radius $R$ is given by
\begin{equation}
\label{eq:force_nonperiodic}
\bm{F}(\bm{r}, R) =  
\left\lbrace
\begin{array}{ll}
 -\bm{r}/R^3, & r < R, \\
 -\bm{r}/r^3, & r > R, \\
\end{array}\right.
\end{equation}
where $r = |\bm{r}|$. To incorporate periodicity, we apply the Ewald summation method \citep{hernquist1991application}. For simplicity, we only account for the contribution from replicas, as in \cite{racz2021anisotropy}
\begin{equation}
\label{eq:force_periodic}
    \tilde{\bm{F}}(\bm{r}, R) = \sum_{\bm{n}} \bm{F}(\bm{r}+\bm{n}, R),
\end{equation}
where a unit simulation box is assumed, $\bm{n}$ represents integer triplets, and $|\left(\bm{r} + \bm{n}\right)| < 2.6$ ensures that the force decays to zero between replicas.

\begin{figure}
\centering
\includegraphics[width=1.0\hsize]{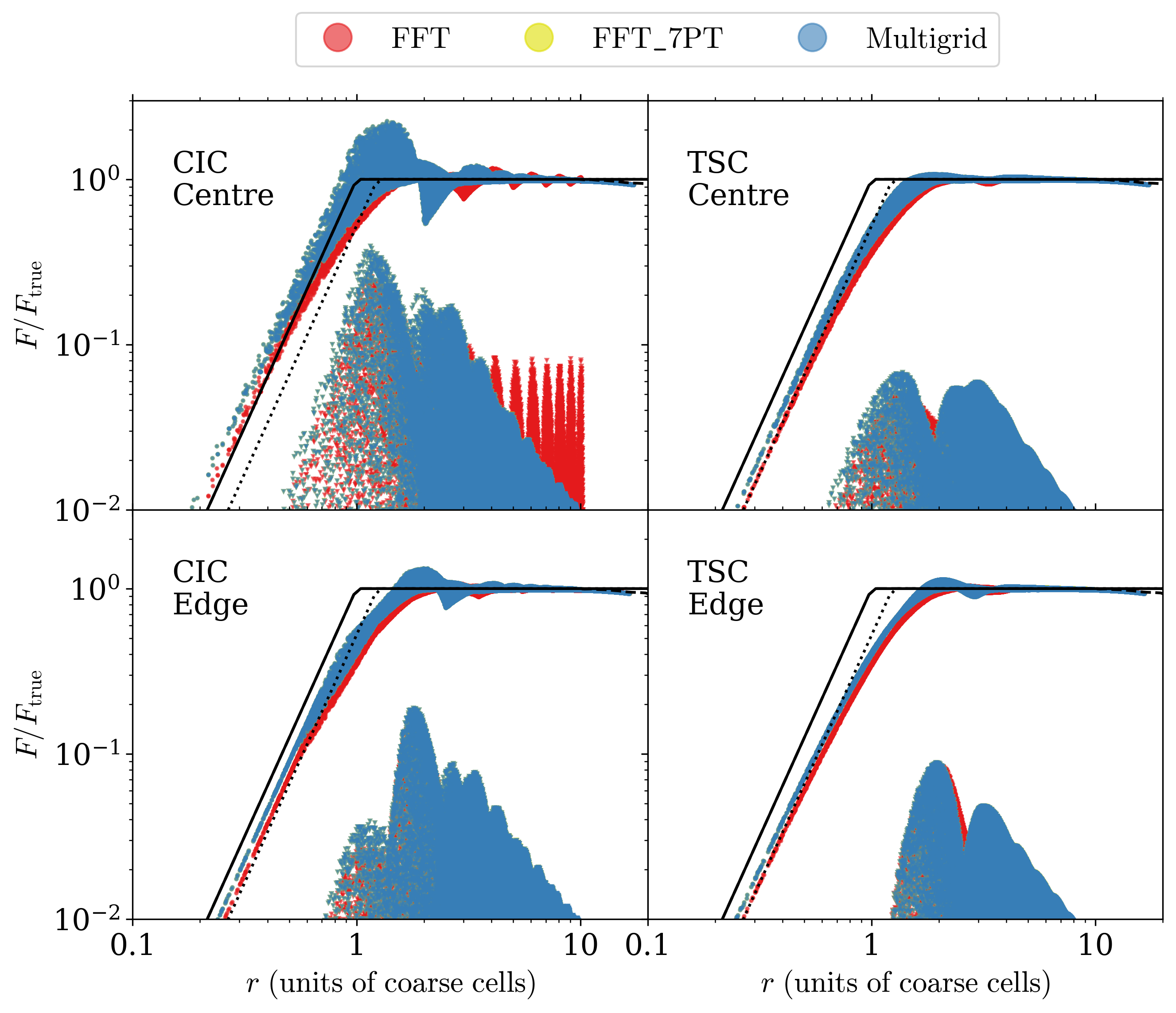}
\caption{
Acceleration of massless particles randomly sampled around a point mass, as a function of distance. The left and right panels illustrate results using the CIC and TSC mass interpolation (and inverse interpolation) schemes, respectively. The top panels show cases where the point mass is positioned at the centre of a cell, while the bottom panels depict cases where it is placed at the edge of a cell.
The red, yellow, and blue data points represent accelerations computed using the FFT solver (uncorrected for the mass-assignment scheme), the FFT\_7pt solver, and the multigrid solver, respectively. Circles denote radial acceleration, while triangles indicate tangential acceleration. 
The black solid line represents the theoretical prediction $\bm{F}(\bm{r}, h)$ from Eq.~\eqref{eq:force_nonperiodic}. For comparison, dotted lines correspond to $\bm{F}(\bm{r}, r_s)$, where $r_s = \left(6/\pi\right)^{1/3} h$ is the radius of a sphere with the same volume as a cell of size 2$h$. Dashed lines illustrate the periodic case $\tilde{\bm{F}}(\bm{r}, h)$.
}
\label{fig:force_point_mass}
\end{figure}

Fig.~\ref{fig:force_point_mass} illustrates the acceleration of $128^3$ massless particles surrounding a point mass in a simulation with $64^3$ cells. 
 The radial accelerations closely follow the predictions for the inner sphere at scales below the coarse cell size $h$. At larger scales, the acceleration transitions to the standard 1/$r^2$ behaviour, albeit with damping due to periodicity. This damping is well captured by Eq.~\eqref{eq:force_periodic}. Notably, the TSC interpolation scheme demonstrates significantly improved isotropy compared to CIC, as evidenced by lower tangential acceleration values.
For the CIC scheme, when the point mass is positioned at a cell centre, only one cell exhibits a non-zero density with a value of $\rho_0$. However, for a point mass located at a cell edge, eight cells have non-zero densities, each with a value of $\rho_0/8$ for CIC and TSC. The sole distinction in this configuration arises from the inverse interpolation from the grid back to the particles. In contrast, for the TSC scheme with the point mass placed at the cell centre, 27 cells exhibit non-zero densities.
In terms of solver performance, the results from the FFT\_7pt and multigrid solvers are consistent and identical, as expected. While the accelerations align at larger distances, the multigrid and FFT\_7pt solvers exhibit a slight overestimation of acceleration at small scales. Put differently, these solvers generate a deeper potential compared to the FFT solver. This increased acceleration results in higher particle velocities, which in turn suppresses clustering at small scales. This behaviour is consistent with the observed damping of the matter power spectrum, as shown in Fig.~\ref{fig:pk_solvers_ramses_amr_tsc}.

\section{Data locality}
\label{appendix:data_locality}
\begin{figure}
\centering
\includegraphics[width=1.0\hsize]{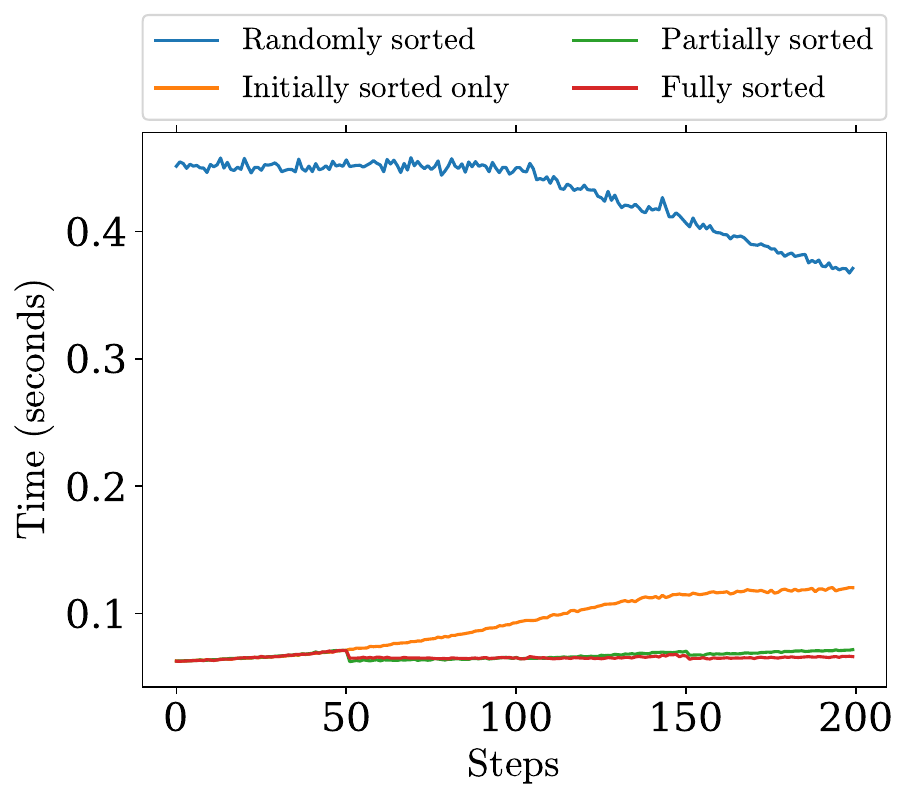}
\caption{Runtime of the inverse TSC algorithm (from mesh to particles) for a force field. In coloured lines we show the results with different ordering of the particle arrays. `Randomly sorted' means that in the initial conditions we randomly shuffle the particle positions. `Initially sorted only' means that in the initial conditions the particles are linearly ordered, with the same contiguous properties as the cells they are in. `Partially' and `Fully sorted' means than every $N$ steps (here 50), we sort the particle array according to Morton indexing based on their positions (see Section~\ref{sec:data_structure}). The simulation has $512^3$ particles and as many cells, and was run with 64 CPUs.}
\label{fig:invtsc_ordering}
\end{figure}

Finally, we highlight the importance of data locality in particle-mesh interactions, focusing on the impact it has on runtime performance. Fig.~\ref{fig:invtsc_ordering} illustrates the runtime for the inverse TSC algorithm under different particle array orderings.
When the particle array is randomly sorted, the inverse TSC takes approximately 0.45 seconds for the first 100 steps, with a gradual decrease down to $\sim$0.37 seconds. In a randomly sorted array, consecutive particles are unlikely to be spatially close, leading to a higher number of cache misses because grid elements are also distant in memory. However, as particles cluster at later times, even a randomly sorted array experiences improved performance due to the higher likelihood that two neighbouring particles are closer spatially, reducing cache misses.

In contrast, if the particle array is initially sorted, the runtime for inverse TSC starts much lower at $\sim$0.04 seconds and increases up to $\sim$0.11 seconds over time. This improvement comes from neighbouring particles in the array being spatially closer at the beginning, allowing the cache to be used more efficiently. The ten-fold improvement in performance, achieved solely by optimising data locality, is a significant gain without requiring complex programming techniques. Over time, as particles move, the initial sorting deteriorates, leading to less efficient cache usage, even though clustering increases.

To maintain performance, \pysco offers the ability to re-sort the particles every $N$ steps (default $N=50$, adjustable by the user). With periodic sorting, the `fully sorted' runtime remains flat as a function of time. However, the sorting operation can become a bottleneck, as \numba does not support parallel sorting algorithms, and resorting to sequential sorting is prohibitive for large particle counts.
To address this, \pysco implements a parallelised, partial sorting method. The particle array is divided into chunks, with $N_{\rm chunks} = N_{\rm CPUs}$, where each chunk is sorted independently by each CPU. The results of this partial sorting are comparable to those of fully sorted arrays, with a slight runtime increase at late times due to the partial nature of the sorting. This increase is marginal, even when using up to 64 CPUs. Therefore, \pysco uses this partial sorting method by default as a balance between performance and computational overhead, especially for large simulations.

\end{document}